\documentclass[acmsmall,nonacm]{acmart}

\acmJournal{CSUR}
\usepackage{graphics}
\usepackage{graphicx}
\graphicspath{ {figures/} }
\usepackage{wrapfig}
\usepackage{xcolor}
\usepackage[]{algorithm2e}
\usepackage{listings}
\usepackage{lscape}
\usepackage{float}

   \definecolor{linkcol}{rgb}{0,0,0.5}
   \definecolor{citecol}{rgb}{0,0.5,0.3}
   \definecolor{urlcol}{rgb}{0.3,0,0}
\usepackage{url}
\usepackage{hyperref}
\usepackage{todonotes}
\usepackage{amsfonts}
\usepackage{amstext}
\usepackage{multirow}
\usepackage[font=footnotesize]{subfig}
\usepackage{fixltx2e}
\usepackage[keeplastbox]{flushend}

\newcommand{\etal}{\textit{et al}.}
\newcommand{\eg}{e.g.,~}
\newcommand{\ie}{i.e.,~}
\newcommand{\etc}{etc.}

\begin{document}
\title{Bridging Information Security and Environmental Criminology Research to Better Mitigate Cybercrime}

\author{Colin C. Ife}
\email{colin.ife@ucl.ac.uk}

\author{Toby Davies}
\email{toby.davies@ucl.ac.uk}

\author{Steven J. Murdoch}
\email{s.murdoch@ucl.ac.uk,}
\affiliation{University College London}

\author{Gianluca Stringhini}
\email{gian@bu.edu,}
\affiliation{Boston University}


\begin{abstract}
Cybercrime is a complex phenomenon that spans both technical and human aspects.
As such, two disjoint areas have been studying the problem from separate angles: the information security community and the environmental criminology one.
Despite the large body of work produced by these communities in the past years, the two research efforts have largely remained disjoint, with researchers on one side not benefitting from the advancements proposed by the other.
In this paper, we argue that it would be beneficial for the information security community to look at the theories and systematic frameworks developed in environmental criminology to develop better mitigations against cybercrime.
To this end, we provide an overview of the research from environmental criminology and how it has been applied to cybercrime.
We then survey some of the research proposed in the information security domain, drawing explicit parallels between the proposed mitigations and environmental criminology theories, and presenting some examples of new mitigations against cybercrime.
Finally, we discuss the concept of cyberplaces and propose a framework in order to define them.
We discuss this as a potential research direction, taking into account both fields of research, in the hope of broadening interdisciplinary efforts in cybercrime research.
\end{abstract}


\maketitle
\thispagestyle{empty}


\section{Introduction}

Society and digital technology have become inseparable.
The most developed countries are on the verge of true digitisation: the Internet of Things (IoT), driverless vehicles, and smart cities~\cite{zanella2014internet}, while even in the poorest of societies, mobile technologies are becoming ubiquitous~\cite{aker2010mobile}.

The Internet (or \emph{cyberspace}) has been described as a `real virtuality'~\cite{castells2002internet}: an interactional environment that is rooted in the real world but transcends its spatial and temporal restrictions~\cite{yar_novelty_2005,llinares_cybercrime_nodate,tranos_death_2013,devriendt_cyberplace_2008,wellman_computer_2001}.
As a result, having a ``digital identity'', ``going online'', or ``surfing the web'' are no longer mere adages, but common, everyday realities.
Now, communities and social networks can be globalised, giving us the ability to communicate in real-time and to meet in this virtual world.
Indeed, our relationships, our activities, and our information are held in ``cyberplaces'', and not just cyberspace~\cite{wellman_physical_2001}.

However, this entrenched influence that digital information now wields over society has also created new opportunities for the cybercriminal economy, which has already proven to be transnational, organised, and incredibly innovative.
From the factory of spam and phishing e-mails~\cite{levchenko_click_2011,stone-gross_underground_2011,mariconti_whats_2017,leukfeldt_comparing_2015,lynch_identity_2005} to meticulously planned romance scams~\cite{buchanan_online_2014,whitty_online_2012,edwards_geography_2018,huang2015quit} and advanced-fee fraud~\cite{mba_flipping_2017,herley_why_2012}, identity theft, cyber fraud, and financial crimes are just some of the profitable avenues for the aspiring cybercriminal.
Anonymous marketplaces and underground chatrooms are diversifying~\cite{dolliver2015evaluating,soska2015measuring}, facilitating the trade of illegal goods and services (drugs, weapons, child sexual abuse images, \etc) with the added benefits and protection of cryptocurrencies and escrow services~\cite{brown2016cryptocurrency,meiklejohn2013fistful,horton2017hard}.
Perhaps most devastatingly, such services have enabled the cybercrime economy to become increasingly organised, with cybercriminals regularly trading services with each other~\cite{sood_crimeware-as--servicesurvey_2013,christin2012traveling,moore_nobody_2010,stringhini_harvester_2014}.
The malware economy is just one manifestation of these rapid developments: growing from a ``cottage industry'' with a few, highly skilled individuals, to a massive and well-oiled criminal business, and constantly being refined, with wave after wave of new distribution vectors and attack patterns.

The interest in cybercrime prevention is profound.
The information security field, which naturally has a strong technical focus, has been studying cybercrime and devising countermeasures since its inception.
On the other hand, the environmental criminology field, which is multidisciplinary at heart and focused on traditional crime prevention based, has been comparatively slow in its reaction to cybercrime.
There have been successes in crime prevention in these respective fields, but keeping up with cybercriminals has proven to be an arms race.
Furthermore, little collaborative or interdisciplinary work between these two fields has been carried out.
To keep up with cybercriminals, there is a pressing need for greater coordination and collaboration amongst computer security researchers and criminologists, among others, to better mitigate cybercrime.

In this paper, we argue that combining contributions from information security and environmental criminology would greatly benefit cybercrime research, both to better systematise past research efforts, and to identify promising future directions that draw from literature in both fields.
To this end, we conduct a literature review of cybercrime literature from the perspectives of information security and environmental criminology, drawing parallels between these two distinct fields and eliciting how theories and frameworks from one map to research in the other.
In this juxtaposition, we identify examples and opportunities for new cybercrime interventions by applying theoretical models from environmental criminology to information security research.
Finally, in arguing the need for a \textbf{new and complementary research direction} that combines contributions from information security and environmental criminology to mitigate cybercrime more effectively, we initiate this process in earnest: we discuss the concept of `place' (amongst other core concepts relating to physical crime) and how we may define the analogous concept of `cyberplace' for cybercrime.
Ultimately, this would aid the transfer and adaptation of crime prevention frameworks to the context of cybercrime so as to provide a new outlook towards fighting it.
We hope that future thought and collaboration in this area would help the device of more effective cybercrime prevention strategies.

In summary, this paper makes the following contributions:
\begin{itemize}
 \item We provide an overview of environmental criminology research, and of how the concepts presented in this area have been applied to cybercrime.
 \item We survey cybercrime research from computer scientists, drawing parallels between the proposed mitigations and well established environmental criminology paradigms.
 To the best of our knowledge, we are the first ones to draw from a wide range of literature and make these parallels explicit.
 \item We set the groundwork for future research directions that could see fruitful collaborations between the two areas.
 To this end, first, we propose some new, potential cybercrime countermeasures using environmental criminology.
 This includes a framework for disrupting malware delivery and botnet operations using Situational crime prevention.
 Again, to the best of our knowledge, no such frameworks have ever before been propositioned.
 \item Second, we reflect on what `place' means for cybercrime (amongst other fundamental concepts  such as space-time, offender behaviours, and  guardianship) and propose a new conceptualisation of `cyberplace,' which combines three fundamental components: \textit{location}, \textit{state}, and \textit{function}.
 We then present some motivating examples of how this concept could be used to derive new methods of cybercrime analysis and mitigations, including the facilitation of environmental criminology techniques for cybercrime.
 We argue that this concept could be applied and developed 
 to identify cyberplaces (websites, services, software) that are at an elevated risk of attracting cybercriminal activity, and, thus, help them to better focus mitigative strategies towards them.
 We are the first to define `cyberplace' in this way for (but not limited to) the context of cybercrime.
\end{itemize}

The rest of the paper is structured as follows.
In Section~\ref{ref:background_place}, we review the evolution of theories and practices in environmental criminology literature, and how they may apply to cybercrime.
In Section~\ref{sec:principles_env_crim}, we evaluate the core concepts of environmental criminology, and how they translate to cyberspace and cybercrime.
In Section~\ref{ref:infosec}, we present an overview of the cybercrime research landscape from the information security perspective.
We then evaluate their mitigations against cybercrime while highlighting some similarities in their approaches with environmental criminology practices.
We end this section by presenting some new, potential mitigations against cybercrime using environmental criminology.
In Section~\ref{sec:cybercrime_place}, we propose a new conceptualisation of `cyberplace' by (1) using an inductive approach to establish examples of `place' contexts through a survey of cybercrimes (Section~\ref{sec:cyber_enabled_dependent}), and (2) considering what `place' means in the real world (Section~\ref{sec:defining_cyberplace}).
We then present some examples of how cyberplaces may later be classified in order to better analyse and mitigate cybercrime (Section~\ref{sec:cyberplace_practice}).
While the majority of this paper is a review of cybercrime literature, in Section~\ref{ref:related}, we provide an overview of previous works on environmental criminology and cybercrime, and `place' concepts in cyberspace.
Finally, we end with concluding remarks in Section~\ref{sec:conclusion}.

\section{The Evolution of Place-Based Theories and Practices in Environmental Criminology }
\label{ref:background_place}
In this section, we first explain why environmental criminology is an appropriate field of choice in furthering our approach to mitigating cybercrime.
We then assess the concept of `place' in environmental criminology theories and practices, how its role has evolved over time, and we juxtapose the applications of these theories and practices between physical crime and digital crime.

\subsection{Why Environmental Criminology?}
\label{why_env_crim}

Before exploring the connections between environmental criminology, information security, and their perspectives on cybercrime, one may be considering at this point, ``why environmental criminology?''
That is, why should we consider this approach in dealing with cybercrime, in association with current information security efforts, and why not, for instance, other criminology subfields, or criminology as a whole?

Criminology is a broad and multidisciplinary subfield of sociology, drawing primarily upon the research of sociologists, philosophers, psychologists, social anthropologists, biologists, and scholars of law.
Criminology possesses an equally broad variety of theories towards understanding crime.
Classical criminology emphasises on the sociological, anthropological, and biological factors affecting one's propensity towards committing crime.
On the other hand, environmental criminology is a unique subfield in that, besides its application of the scientific method to examine crime, it draws focus on the (previously overlooked) environmental and circumstantial factors that create criminal opportunity, rather than purely focusing on the individual characteristics alluding to the ``criminal profile.''
In this regard, environmental criminology draws on research from a broader range of technical fields, from geography and economics to computer science and mathematics, to focus on and assess the proximal (rather than distal) aspects of a crime event that explain why it occurs, thus pointing to how it could be deterred.
Environmental criminology, therefore, manages to elucidate how and why crime is not bounded to only those who ``fit the criminal profile,'' but can be committed by any member of society, and why one may be found to commit a crime that others would deem contrary to their disposition or ``character.''
Crime science~\cite{cockbain2017crime} is an evolved field of environmental criminology with the principle focus on controlling crime and reducing ``harm.''
Because of this relationship between the two fields, we will generally focus on the role of environmental criminology for the remainder of this paper, considering crime science as its natural symbiote.

Returning to criminology, and looking at the dimension of efficiency, classical criminology approaches lead to a heavy reliance on the judicial and penal systems to inhibit crime, through deterrence, punishment, and rehabilitation.
Though such are likely necessary for society, they are, nonetheless, flawed as a sole solution.
First, there is the attrition of justice: ever-diminishing proportions of offenders are successfully reported, then arrested, then indicted, then imprisoned, and then rehabilitated~\cite{garside2004crime}.
Therefore, the majority of offenders remain within or are
 reintroduced into wider society with little to no lasting, positive change.
However, environmental criminology takes a pragmatic approach from the outset. With the primary focus on pre-empting and preventing crimes before they occur, this approach favours manipulating the immediate environment so as to deter one from committing a crime, such as by increasing the perceived risks or costs, reducing the perceived rewards or provocations, or removing the excuses associated with the commission of a crime~\cite{clarke1997situational}.
In essence, this approach aims to reduce criminal opportunity for potential offenders.

Finally, the environmental criminological approach favours direct and practical methods, techniques, and technologies over purely theoretical discourse.
Improvements in modern technology have resulted in the development of a number of crime analysis techniques (\eg crime scripting~\cite{cornish1994procedural}, agent-based modelling~\cite{bonabeau2002agent}, geographic profiling~\cite{rossmo1999geographic}), crime prevention techniques (\eg situational crime prevention~\cite{clarke1997situational}, crime prevention through environmental/urban design~\cite{jeffery1977crime,newman1972crime}), and tools (\eg digitised crime mapping systems and hot spot policing~\cite{eck1995crime}).
With this constant evolution alongside modern technology, it is no wonder that environmental criminologists have begun to shift their focus crimes committed using computer systems and the Internet.
Given the similar focus held by information security researchers on mitigating cybercrime, it is equally unsurprising that these two fields could serve to complement each other in achieving this shared goal.

\subsubsection*{Criticisms and Challenges}

The field of environmental criminology does not come without its critics.
One common and often argued criticism of environmental criminology is that crime interventions derived from it do not lead to an overall reduction in crime -- they only cause crime to move elsewhere.
This side-effect of crime intervention is a phenomenon known as \textit{displacement}.
However, most displacement research has found that displacement is far from an inevitable side-effect of crime intervention.
Instead, crime interventions have been generally found to deliver net benefits with reduced crime~\cite{johnson2014crime}.

More generally and as summarised by Wortley and Tilley~\cite{Wortley2014}, one of the main criticisms of environmental criminology is that it ignores the ``root causes'' of crime.
As such, prevention efforts based on this approach are often characterised as only catching the ``low-hanging fruit.''
Wortley and Tilley argue (as we will find in our review of the relevant theories and practices of environmental criminology) that human behaviour -- and hence criminal behaviour -- are by their very nature situational and intricately linked with the immediate environment.
As such, the immediate situational and environmental aspects of a crime event are root causes of the crime themselves.
With that being said, the proposition of environmental criminology is not to be a comprehensive solution to all crime -- crime in itself is a highly complex phenomenon.
Rather, as in its designation, it is designed to tackle the immediate environmental aspects that can lead to crime so as to control it.

With that being said, one may also consider what benefit is there in investing in interdisciplinary research efforts between two fields that already share a similar focus (information security and environmental criminology).
As we will see in later sections, the main benefit of a unified approach between the two fields, we believe, is the added structure and systematisation to the processes of devising, implementing, and monitoring cybersecurity mitigations.
In essence, our argument is that the crime prevention theories and practices of environmental criminology may be extended to supplement the techniques already applied in information security.

\begin{figure}[t]
    \centering
    \includegraphics[width=0.6\textwidth]{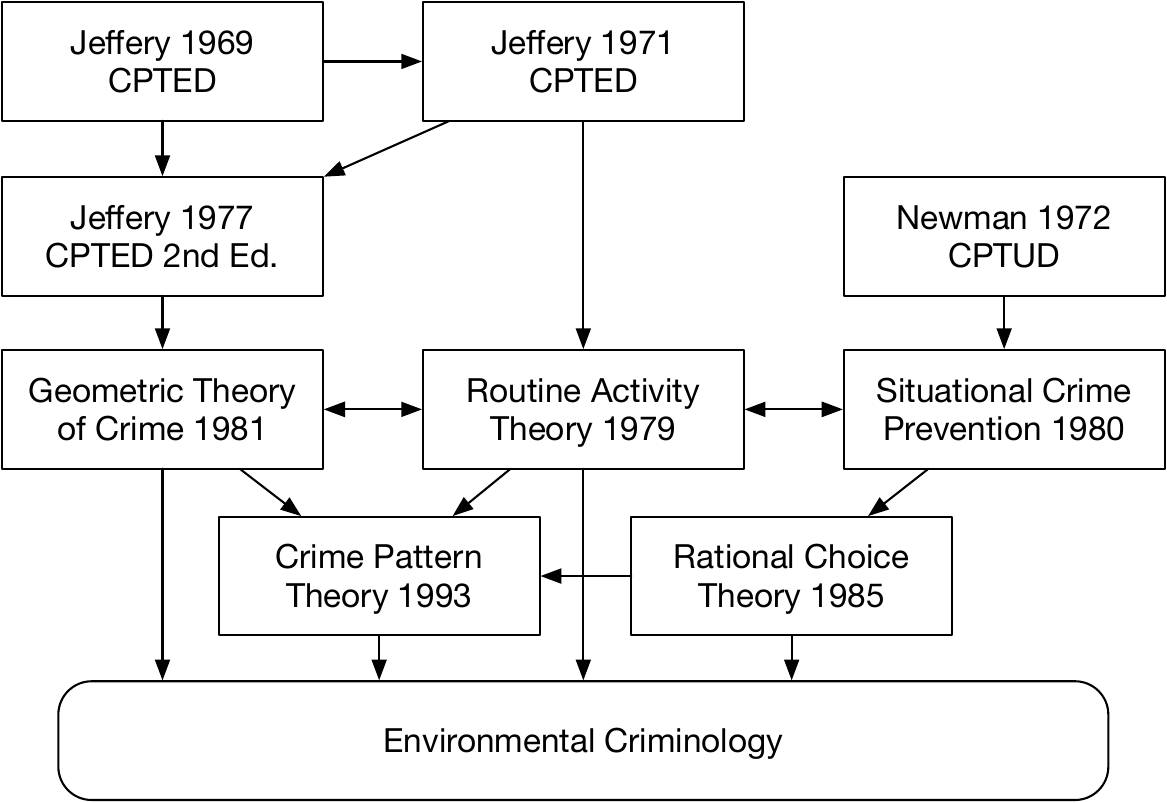}
    \caption{The evolution of environmental criminology.
    }
    \label{fig:evolution}
\end{figure}

\subsection{Theories within Environmental Criminology} \label{sec:theory}
The foundation of environmental criminology (\ie the study of crime, criminality, and victimisation) has a primary emphasis on the specific places and times where and when crime events occur.
In environmental criminology, it is theorised that the characteristics of the immediate environment have a significant effect on whether a potential offender commits a crime or not.
These environmental features are greater emphasised than other (distal) factors that are typically proposed within classical criminology theory, such as the anthropological or neurological characteristics of ``deviant'' offenders.

We provide a condensed background of the concept of `place' in environmental criminology, and how its role within this field has evolved over time.
Figure~\ref{fig:evolution} summarises this evolution of environmental criminology.

\subsubsection{Early Studies}
The idea that crime is non-uniformly distributed in space is not new.
In fact, criminology research on these premises stretches back to almost 200 years, ranging from studies by Guerry~\cite{guerry1833essay}, Quetelet~\cite{quetelet1842treatise}, and Glyde~\cite{glyde1856localities} on early crime mapping and statistical applications to the social sciences, to works by Burgess~\cite{burgess1916juvenile} and Shaw and McKay~\cite{shaw1942juvenile} on the links between juvenile delinquency in urban areas and social disorganisation theory.
A common thread within these works is the recognition of the fact that, with regards to areas and locations, crime does not occur uniformly.
Instead, crime is heterogeneous in space.

Moving forward to the 1970s and 80s, environmental criminology called for a shift in focus on the specific
places where crimes occur over other situational factors, such as the motivations of the offender, or the exploitability of the victims and/or targets of crime.
Simply put, the ``where'' of crime was considered as or more important than the ``who'' or the ``what'' of crime.
This led to the emergence of several fundamental environmental theories to explain crime, which, we believe, are better suited to mitigate new forms of crime, such as cybercrime.

\subsubsection{Crime Prevention Through Environmental Design.}
Jeffery~\cite{jefferyray} coined and formulated this term, often abbreviated as CPTED.
The CPTED concept is simple: just as buildings and properties are designed to prevent damage from the forces of the elements, they should also be designed to deter and prevent crime.
Such techniques include, for example, using a single, clearly notable point of entry to a private property to enable easy access control, or making communal areas highly visible to enable natural surveillance by its residents.
Around the same time, Newman~\cite{newman1972crime} developed Crime Prevention Through Urban Design (CPTUD) and the concept of \emph{defensible space}: a model of residential environments that exhibit territorial behaviours and senses of community to deter criminal activity within them.
Ultimately, this model aims to provide perceptible cues to potential offenders that these areas are defensible (clearly bounded, regularly monitored, limited escape routes, territorial residents and users, \etc) and, thus, unconducive to crime. Some argue that CPTUD could be interpreted as an application of CPTED through a subset of its dimensions
~\cite{jeffery1977crime,andresen_environmental_2014}.

There are six key principles of CPTED, which, in varying degrees, are also relevant to cybercrime:
\begin{enumerate}
    \item \textbf{\emph{Territoriality:}} people tend to lay claim over an area that they have some form of ownership and will defend them against intrusion.
    Similarly, in the cyber context, people employ various methods of control over their computer systems, accounts, and websites to prevent unauthorised access (\eg the use of passwords and security protocols) or to prevent malicious behaviours within them by enforcing terms of conditions for their users.

    \item \textbf{\emph{Surveillance}:} architectural designs that encourage residents to inhabit and interact with public spaces are more likely to deter criminal behaviour, such as by implementing increased lighting and unobstructed lines of sight.
    The concept of natural surveillance (as well as collective guardianship), is also useful in cyberspace.
    Particularly within social networks, forums, and e-commerce websites, members of these services can flag inappropriate or illegal content (and their users) for removal and can report software bugs in these services.

    \item \textbf{\emph{Target hardening:}} one may implement physical barriers (fences, gates, locks) to inhibit forced entry.
    In a digital sense, this maps to the use of authentication (password-protection, cryptography) and security technologies (antivirus, antimalware, firewalls, intrusion detection/prevention systems), and security personnel (network administrators, security analysts), which increases the difficulty for threat actors to compromise a service, system, or network.

    \item \textbf{\emph{Access control:}} one may deter criminal activity by defining site boundaries (fences, hedges), limiting access to a single point of entry or exit, implementing security systems and personnel, or guiding movement through a site.
    In cyberspace, this could involve the use of various authentication and security technologies as in target hardening, or utilising control flow integrity~\cite{abadi2009control} and user experience (UX) design techniques.

    \item \textbf{\emph{Maintenance:}} a well-maintained site sends a signal to outsiders that people notice and care about what happens in the area, while those that are not well-maintained are more likely to entice vandalism, and, consequently, higher levels of crime (`Broken Windows' theory~\cite{wilson1982broken}).
    This principle is also relevant in the context of cybercrime.
    For instance, websites that are regularly monitored may be less desirable avenues for posting malicious or illegal content.
    Software that is regularly patched may be a more difficult target for attackers to exploit.

    \item \textbf{\emph{Activity support:}} providing clear signage on what are acceptable and unacceptable behaviours within a site can encourage expected patterns of use within it, \eg including `entrance' and `exit' signs, or notices of criminal prosecution against malefactors.
    Likewise, shared services and websites that enforce terms and conditions (no hate speech, spam, or malicious hyperlinks, \etc) are likely to encourage compliant behaviour (at least from real users) and discourage inappropriate ones.

\end{enumerate}

CPTE/UD are a subset of the more general \emph{Design Against Crime} (DAC) principles, as proposed by Poyner~\cite{poyner1983design}, which also includes crime prevention through product design~\cite{clarke1999hot}.
On reflection, one may find that there are similarities between these crime prevention approaches and the security-focused design and maintenance procedures in systems security, such as the various levels of application security (secure coding, secure operating systems), system security (firewalls, antivirus, and antimalware software), and network security (intrusion detection/prevention systems, security information and event management systems).
However, in CPTE/UD, there is also a clear focus on empowering communities to deter criminal behaviour in their locales.
Though one may argue that the use of computers is much more solitary than interacting in the real world, in actuality, as highlighted by other researchers~\cite{wellman_computer_2001,wellman_physical_2001,sussan2006location}, there is a great degree of online community in various forms, such as the shared use of computers and networks, software applications, forums, social networking, and e-commerce sites.
Thus, these principles could be adapted to harness the power of online communities so as to prevent malicious behaviours within them.

\subsubsection{Situational Crime Prevention.}
Following the works of CPTE/UD~\cite{jefferyray,newman1972crime}, and the successes of their implemented interventions, Clarke~\cite{clarke1980situational} argued for a situational approach to crime prevention.
This approach is premised on offenders being rational actors, and that their choices and decisions towards crime are influenced by the characteristics of their immediate environment.
For instance, a window regularly being left open could influence the commission of a burglary on that property, whereas a visibly secure property would be more likely to deter such a crime.
Over time, and after some literary discourse~\cite{wortley_classification_2001}, Cornish and Clarke updated the situational crime prevention (SCP) framework to establish 25 techniques~\cite{cornish_opportunities_nodate}.
The SCP framework can be summarised under five categories of techniques to deter potential offenders from initiating a crime event:

\begin{enumerate}
    \item \textbf{\emph{Increase the perceived effort.}} Physically, this includes the use of site security to deter offenders, while for cybercrime, this could involve automatically patching software and utilising application and network firewalls to deter hackers and malware.
    \item \textbf{\emph{Increase the perceived risks.}} Mitigations include implementing CCTV surveillance, or, for cybercrime, employing identification checks for money transfer services or de-anonymising cryptocurrency transactions.
    \item \textbf{\emph{Reduce the anticipated rewards.}} Examples of these mitigations include anti-theft mobile apps that can lock phones remotely. For mitigating cybercrime, this includes using system backup policies to empower ransomware victims against complying to criminal demands, or using digital watermarking to detect piracy.
    \item \textbf{\emph{Reduce the provocations.}} This categority includes crowd control measures at venues to minimise stress and prevent altercations, or, in the digital sense, swiftly detecting and removing abusive and illicit content (which could breed further illegal activity) using automated filters and user reporting procedures.
    \item \textbf{\emph{Remove the excuses for crime.}} Examples include displaying roadside speed signs or using breathalysers in pubs, whereas for cybercrime, mitigations include displaying and enforcing stricter rules for social networking and e-commerce sites to discourage offensive and illegal behaviours.
\end{enumerate}

Though this framework was developed predominantly for urban crime, it has shown to be useful in a wide variety of crime scenarios.
For example, one adaptation has been made for counter-terrorism purposes~\cite{clarke2006outsmarting}, which maps the five types of SCP techniques across four necessary components of terrorism (targets, tools, weapons, facilitating conditions), generating a lattice of potential mitigations.
Besides applying the 25 SCP techniques to various types of cybercrime, future research could go into developing customised frameworks for each type of cybercriminal operation, such as further adapting the counter-terrorism variant of SCP towards mitigating cyberterrorism (hacktivism, denial of service attacks) or malware delivery operations.

There are two potential effects of an intervention that are also the main criticisms of SCP~\cite{reppetto1976crime,gabor1981crime}: \emph{crime displacement} and \emph{crime adaptation}.
Crime displacement involves the movement of crime (\ie in space, time, modus operandi (MO), crime type, or the perpetrators and/or targets involved) as a direct result of a crime intervention.
Cornish and Clarke~\cite{cornish1987understanding} use rational choice theory to attempt to explain this phenomenon.
Crime adaptation involves offenders learning of an intervention and adapting their techniques or MOs in order to bypass that intervention and commit the same crime.
These crime phenomena could be of particular importance to cybercrime, as the ability of cybercriminals to move their operations elsewhere is effectively ``free,'' in comparison to criminal operations in the real world, which can be more challenging.
For instance, once a server being used for spam or other malicious activities is blacklisted, the operator could just change its IP address or domain name to circumvent this blacklist and continue their operations.
Moreover, though there is a wide variance in their skill sets, some cybercriminals actively seek to beat the best cybersecurity defences and identify new ways to get around them (\eg zero-day exploits, anti-analysis functionalities such as polymorphism and VM detection in malware).
Therefore, it is even more important to devise interventions that are difficult to circumvent, or that would at least reduce the profits or increase the efforts and/or risk for cybercriminals who would do so.

%

\subsubsection{Routine Activity Theory.}
Cohen and Felson~\cite{cohen_social_1979} proposed this theory as a macro-level explanation for crime rate changes in the United States between 1947 and 1974.
This theory states that crime is less affected by (traditionally postulated) social causes, such as poverty, inequality, or unemployment, but more so by the immediate opportunity for one to commit a crime.
In essence, they propose that ``crime follows opportunity.''
That is, as more opportunities for crime arise, more crimes will occur.
The core of this theory postulates that for (direct-contact predatory) crime to occur, three necessary components must physically converge in time and space: (i) a motivated offender, (ii) a suitable target, and (iii) the absence of a capable guardian (or some other controller, such as one who can handle the offender, or a place manager).
In the real world, this could be exemplified by sexual assaults being more common at night as either the offender or the victim (or both) is more likely to be intoxicated and as there are fewer people out in public to deter them~\cite{dedel_2011}.
The same principle seems to apply in the cyber world.
For example, studies have shown that botnet activities peak during the day and drop off at night~\cite{stone2011analysis}, while most malware delivery is carried out on weekends rather than weekdays~\cite{unpublished_ife_snapshot} -- this is in line with when computers are used most often, especially for leisurely activities.
Cyberharassment and cyberbullying can only occur when the victims ``come online'' or access peer-to-peer services (online forums, e-mail services, social network sites, messenger apps).
Hackers and their malware can only infiltrate a target system when it becomes available through a connecting medium (\eg the Internet, a drive-by download on a website, downloading an e-mail attachment, or accessing an infected USB device).

\begin{figure}[t]
    \centering
    \includegraphics[width=0.35\textwidth]{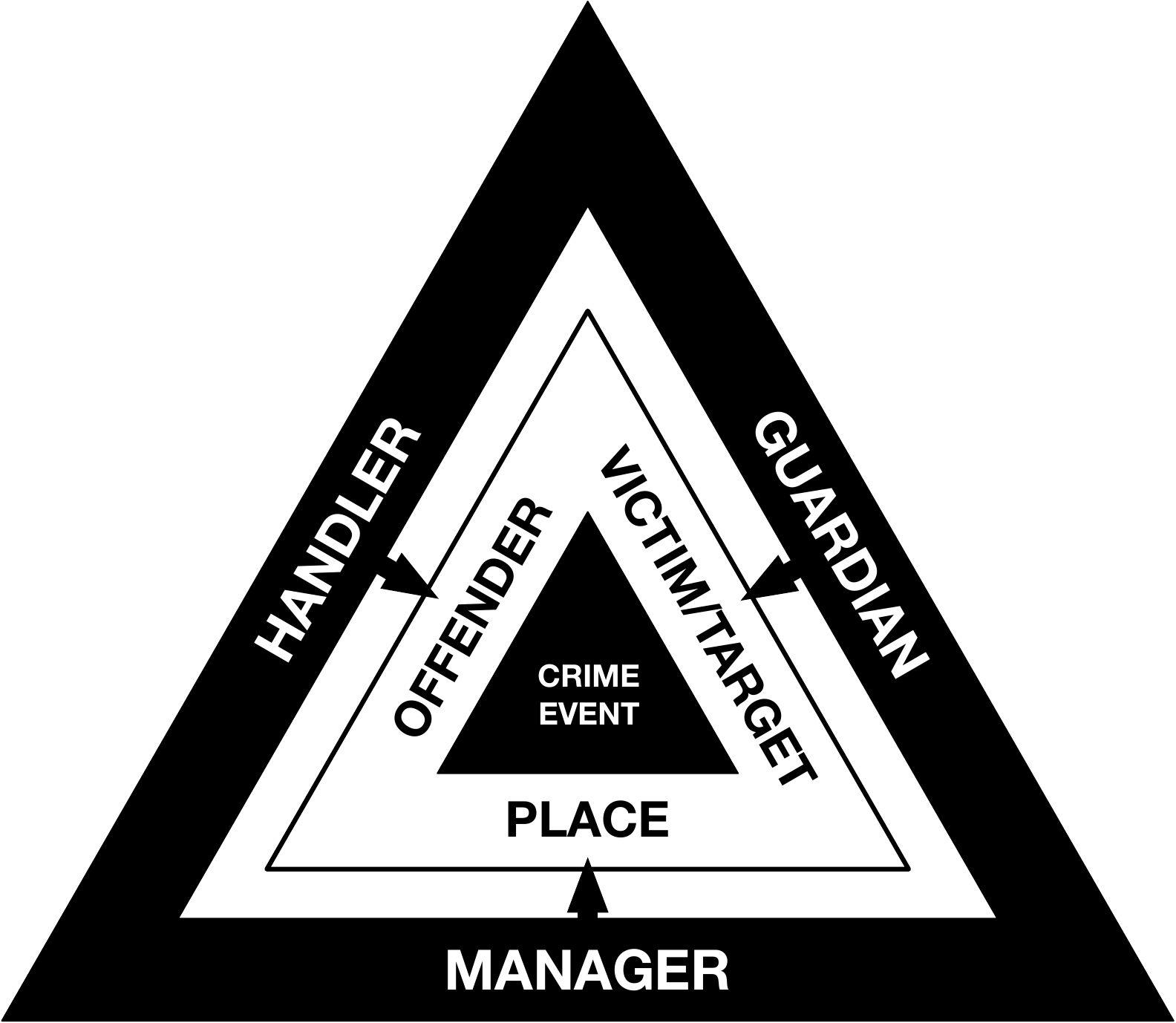}
    \caption{The `crime triangle' of routine activity theory.}
    \label{fig:crime_triangle}
\end{figure}

Over time, this simple but powerful trifactorial relationship has come to be known as the Crime Triangle (Figure~\ref{fig:crime_triangle}), and is a key component of environmental criminology.
Cohen and Felson also identify that most predatory crimes involve rational decision-making by the offender, particularly in qualifying ``suitable'' targets.
With this premise in mind, they introduce the \emph{VIVA} model (Value, Inertia, Visibility, and Accessibility) to explain how offenders qualify and select victims and targets, and how altering such dimensions could affect their perceived suitability for victimisation from the perspectives of these offenders.
Alternative models have also been proposed, such as {CRAVED} (Concealable, Removable, Available, Value, Enjoyable, and Disposable), which is specifically designed for theft targets, or ``hot products''~\cite{clarke1999hot}.
As routine activity theory is core to environmental criminology, we will revisit the fundamental concepts of space and time, offender behaviours, guardianship, and how offenders identify suitable targets, and explore how these concepts apply in cyberspace and within cybercrime (Section~\ref{sec:principles_env_crim}).

\subsubsection{Geometric Theory of Crime.}
Building on routine activity theory, Brantingham and Brantingham~\cite{brantingham1981introduction} focus on the spatio-temporal relationship of crime.
To elaborate, just as crime is non-uniformly distributed in space, it is also non-uniformly distributed in time.
This theory accounts for ``peaks'' and ``troughs'' in criminal activity across different geographical areas and places.
The authors also identify a further three dimensions in understanding crime: the \emph{legal dimension} (the creation, perception, and governing of laws); the \emph{offender dimension} (the motivations of the offender and how they vary in time); and the \emph{victim dimension} (why offenders select certain targets).

Going further, Brantingham and Brantingham~\cite{brantingham1981notes} attempt to account for the non-uniformity and non-randomness of crime in their Geometric Theory of Crime.
This work focuses on the urban landscape, theorising that an offender, just like non-offenders, will spend most of their time engaging in normal routines of non-criminal activity.
It is through these routine activities that an offender develops their ``personal awareness space'', \ie the areas and routes with which they are most familiar.
It is theorised that when these awareness spaces intersect with the activity spaces of victims, it is in these areas that offenders conduct most of their criminal activity.
The authors explain this theory by discretising the real world into
(i) \emph{nodes}: places which are central to the lives of people, and to and from which they travel (\eg shops, schools, workplaces) -- offenders tend to search for opportunities here;
(ii) \emph{paths}: routes that link nodes -- people are often victimised along paths; and
(iii) \emph{edges}: physical and/or perceptual boundaries (\eg rivers, major arterial roads) that separate distinguishably different areas  -- ``outsiders'' tend to commit crimes at these boundaries, while ``insiders'' tend to commit crimes within the bounded areas~\cite{brantingham1993nodes,brantingham_criminality_1995}.

The relationship between this theory and the activities in cyberspace is interesting.
As we discuss further in Section~\ref{sec:space_time}, cyberspace is very different in its construction as compared to the real world: it is highly discretised (as opposed to the contiguity of physical space), with transitions between one online site to another being almost instantaneous.
In this regard, the concept of victimisation occurring more along paths would need revision.
However, it can be understood that online users form their own awareness spaces in cyberspace, based on the websites that they frequent and the services that they use.
In turn, this familiarity may indeed reduce the perceived risks of using such services by these users (such as the continued use of illegal streaming and piracy sites), while also allowing cybercriminals to identify suitable attack vectors (\eg vulnerabilities in a website) and targets (\eg vulnerable users to socially engineer).
This concept of awareness space is also comparable to the `reconnaissance' stage of the Cyber Kill Chain~\cite{hutchins2011intelligence} model, such as when crawler bots scrape websites for e-mail addresses, or when malware scans nearby devices for vulnerabilities.

\subsubsection{Rational Choice Theory.}
For several years, rational decision-making had been a given assumption in modelling offender behaviours (\eg situational crime prevention, routine activity theory).
However, Clarke and Webb~\cite{clarke1985modeling} formally proposed the Rational Choice Theory in 1985 to evaluate this proposition.
By viewing different crime events in terms of (perceived) opportunity, costs, and benefits, this theory provides a possible explanation as to why and how offenders make rational decisions towards committing crime, as well as what may prevent such decisions from being made.
This theory makes sense with physical crime, but perhaps even more so in the case of cybercrime.
That is, there is a considerable rational element that may motivate one to commit a crime in the real world.
However, other non-cognitive factors lead to a crime event, such as the immediate stresses, pressures, or prompting cues~\cite{wortley_classification_2001} that can, for example, spark an altercation in a bar, which may ultimately lead to an aggravated assault or even manslaughter.
But, when we consider cybercrime, how these ``situational precipitators'' affect decision-making is not as clearly understood.
Granted, the commission of a wide range of cybercrimes may require skill, motivation, and careful thought.
For instance, it is probably far-fetched to identify a complex ransomware operation, spanning several weeks of activity, as a ``spur-of-the-moment'' assault.

Nonetheless, there are examples of cybercrimes that may be committed without prior intention, or even the knowledge that they are crimes.
For example, one's participation in a heated online argument, or their reaction to an emotive piece of news, could quickly escalate to online abuse or cyberharassment.
A user may stumble upon a bug in a website, and, instead of reporting it to the site owners, may be tempted to see to what end exploiting that bug may lead, \eg accessing accounts that had just had their credentials leaked in plaintext.
Even hackers, though they may be considered as skilled and rational actors, may not always be aware of the criminality of some of their actions in cyberspace, thus, failing to distinguish between that which is lawful and unlawful.

\subsubsection{Crime Pattern Theory and Repeat Victimisation.} \label{sec:crime_patterns}
The links between routine activity theory, the geometric theory of crime, and rational choice theory are apparent. Brantingham and Brantingham~\cite{brantingham1993environment} attempted to leverage these links and synthesise these theories within Crime Pattern Theory.
Through this theory, criminologists and crime scientists have been able to explain why crime occurs in certain areas, based on the intersecting activity spaces of offenders and their victims or targets.
This theory classifies three types of crime hotspots~\cite{brantingham_criminality_1995}:
\textit{crime attractors}, which are places that are well-known to offenders for illegal activity and abundance in criminal opportunity, such as bars and nightclubs; \textit{crime generators}, which are places that attract large crowds of people and where, being amongst them, offenders become aware of criminal opportunities there, such as shopping malls, schools, and entertainment venues; and \textit{crime enablers}, which are places that facilitate crime due to their lack of place management, such as public parks and parking lots.

Repeat victimisation is a specific type of crime pattern, relating to the heightened risk of a victimised individual, demographic, property, or location, to being victimised again~\cite{farrell1993once}.
For example, the vast majority of homes remain unburgled while a minority of homes suffer multiple burglaries in a year.
This increased risk of victimisation can be understood by the ``flag'' explanation, which relates to the characteristics of the victim or target that make them desirable to offenders (\eg a home with broken locks or overgrown bushes blocking public visibility), or by the ``boost'' explanation, which relates to the role of repeat offenders in these crimes (\eg a burglar identifying when a home is vacant, or learning techniques to overcome a security system)~\cite{pease1998repeat,ashton1998repeat}.
The problem of repeat victimisation is that it is a concentration of a majority of crimes involving only a few victims and targets, and being commissioned by a few offenders.
As is the case with crime displacement, there are a number of factors that could cause this increased concentration of crime, such as geography (\ie crime hotspots), types of (risky) locations (\eg shopping malls, schools), the availability of targets (\ie ``hot products''), or the presence of repeat offenders or chronic victims.
Besides the same people being repeatedly victimised, there are also \emph{near-repeat victims}, who are different victims to the same (or similar) crime with some similarity to the initial victim (\eg houses near to the one that was burgled are more likely to suffer a burglary than those further away).

\noindent There are many examples of crime patterns also occurring in cyberspace (Table~\ref{table:crime_patterns}).

\begin{table}
\centering
\caption{Examples of crime patterns in cyberspace.}
\label{table:crime_patterns}
\resizebox{1\linewidth}{!}{
\begin{tabular}{l|l}
\toprule
\textbf{Crime Pattern} & \textbf{Example} \\
\midrule \midrule
\textit{\textbf{Repeat and Near-Repeat}} & Cyberharassment against a victim and their close contacts~\cite{henry_technology-facilitated_2018,reyns_situational_2010,tokunaga_following_2010}. \\
\textit{\textbf{Victimisation}} & Repeat sale and use of credit cards from Dark markets~\cite{moore_nobody_2010}. \\
& Repeat use of stolen accounts~\cite{onaolapo_what_2016}. \\
& Malware spreading to nearby devices~\cite{moore_code-red:_2002,bose_agent-based_2011}. \\
& Continued botnet activity: PPIs, spam operations, DDoS attacks,\etc~\cite{binsalleeh_analysis_2010,stone2011analysis,stringhini_harvester_2014,stone-gross_underground_2011}. \\
& Malvertisement drive-by downloads on same vulnerable browser plugins~\cite{zarras2014dark,sood_malvertising_2011,ford_analyzing_2009}. \\
& Outdated WordPress sites re-compromised twice as often as sites with up-to-date versions~\cite{vasek_hacking_2016,vasek_identifying_2014}. \\

\midrule \midrule
\textit{\textbf{Crime Generators}} & E-commerce sites that enable buyer and seller fraud. \\
& Online forums and game servers that enable fraud, sexual exploitation, and harassment. \\
& Windows OS, which has 80\% market share and fewer vulnerabilities than others but is the most attacked OS~\cite{noauthor_top_nodate}. \\
& Websites with popular CMSes, which are more likely to be hacked or suffer fraud than others~\cite{soska_automatically_2014,vasek_hacking_2016,vasek_identifying_2014}. \\

\textit{\textbf{Crime Attractors}} & Underground marketplaces such as Silk Road, which facilitate illegal products and services solicitation. \\
& P2P software, pirate websites, and illegal streaming services are at an increased risk of delivering malware~\cite{telang2018does}. \\
& Browser extensions hosting malvertisements~\cite{zarras2014dark,sood_malvertising_2011,ford_analyzing_2009} and leading to other malware intrusions~\cite{kotzias2016measuring,thomas2016investigating}. \\
& Illegitimate app stores: large parts of the world can only access these though at higher risk to mobile malware~\cite{ng2014android}. \\

\textit{\textbf{Crime Enablers}} & Social media sites with limited parental or speech-detecting guardianship enabling cyberharassment. \\
& Websites and software with well-known vulnerabilities (SQLi, XSS, \etc) and not regularly monitored or updated. \\
& Devices and networks with no antivirus software or firewall protection attractive to malware. \\
& Internet service providers with poor security hygiene attract malicious websites~\cite{stone2009fire}. \\
& Criminals heavily use banks with poor operational security to handle their fraudulent payments~\cite{levchenko_click_2011}. \\

\bottomrule
\end{tabular}
}
\end{table}

\subsection{Practices of Environmental Criminology}


We have already discussed some theoretical models and how they can be used to analyse various crime types.
In this section, we will cover some of the practical applications of these theories.

\noindent \textbf{Action research models.} There are several systematic processes and risk management frameworks that have been used to implement crime prevention in a variety of public and private contexts (\eg SARA~\cite{eck1987problem}, the 5Is~\cite{ekblom2010crime}, ISO 31000\footnote{\url{https://www.iso.org/iso-31000-risk-management.html}}).
However, as other researchers~\cite{clarke1997situational,haelterman2013supply} describe, the common thread between these different approaches is that they are action research models, allowing researchers and practitioners to work together to:
\begin{enumerate}
    \item analyse and define the problem, the ecosystem, and the relevant stakeholders (\eg a shared computer, its programs, and its users),
    \item analyse the situational conditions that permit or facilitate the crime event under study (\eg infected app, malvertisement/drive-by, or socially engineered download),
    \item identify, evaluate, and implement potential countermeasures (\eg block third-party/untrusted apps by default, regular software updates, e-mail ``safe links'', online safety reminders), and
    \item assess the effects of these measures (\eg diagnostics, user feedback), reiterating as necessary.
\end{enumerate}

We can identify system vulnerabilities (such as for an OS environment, or a sociotechnical system) using the risk management variant of this approach, chiefly by identifying the parameters of a ``good'' system state, the goals and assets of the relevant stakeholders, and triggering events (criminal or mistaken) that counteract these goals as deviations from this state.
Islam~\etal~\cite{islam2019socio} propose such a framework for reducing human-related risks in sociotechnical systems.

Table~\ref{table:crime_science_examples} shows some examples of action research applied to real crime problems.

\noindent \textbf{Hot spot policing}~\cite{eck1995crime}, which is based on crime pattern theory, involves constantly developing models of crime ``hotspots'' (points, streets, areas, chronic victims) and focusing law enforcement resources within them to efficiently deter crime.
This approach makes sense, as focusing limited resources on the biggest sources of crime is likely to reap the most benefit overall.
As we discuss later, this could also be applied to focusing resources on cyberplaces (websites, services, applications, \etc) at an elevated risk of malicious activity (see Section~\ref{sec:cybercrime_place}).

\noindent \textbf{Geographic profiling}~\cite{rossmo1999geographic} is an investigative technique to locate a serious offender's ``anchor point'' (\eg their home or workplace), by connecting the locations of a series of crime events.
This approach is similar to the clustering techniques developed by researchers~\cite{meiklejohn2013fistful,harlev2018breaking} that can (to some extent) de-anonymise the operators of Bitcoin transactions involving illegal activity, or the techniques that can be used to carry out traffic correlation~\cite{murdoch2005low} and de-anonymise Tor\footnote{\url{https://www.torproject.org/}} users.

\noindent \textbf{Crime scripting}~\cite{cornish1994procedural} is an analytical technique that is used to extrapolate the sequence of steps an offender may take to commit a criminal offence.
For example, in a romance scam, fraudsters create a fake account on a dating service, they identify a suitable victim,
they go through a grooming phase, followed by the actual fraud when the scammer asks their victim for
money.
Dissecting the various steps of an offence can be useful to better understand it and to identify
potential interventions.
The Cyber Kill Chain~\cite{hutchins2011intelligence} is a crime script example that is already used for analysing system intrusions.

\noindent \textbf{Agent-based modelling}~\cite{bonabeau2002agent} (ABM) is a class of computational models that can be used to simulate an environment and assess how different actors interact with it and with each other.
It is an analytical technique that is widely used in biology, sociology, computer science, and criminology.
Researchers~\cite{bose_agent-based_2011} have used ABM to model malware activities in heterogeneous environments.


\begin{table}
\centering
\caption{Examples of environmental criminology applied to crime problems.}
\label{table:crime_science_examples}
\resizebox{1\linewidth}{!}{
\begin{tabular}{l|l|l|l}
\toprule
\textbf{Crime Type} & \textbf{Theoretical Model/s} & \textbf{Intervention/Implication} & \textbf{Cybercrime Analogue/s} \\
\midrule \midrule

Number plate  & Design against crime, VIVA, & \textbf{Anti-theft plate~\cite{doc_case_studies_2015}}: breaks upon removal (\textbf{\emph{inertia/removable}}), rendering it useless  & Spoofing attacks, online  \\
theft & CRAVED & for further criminal activities (\textbf{\emph{value}}). & identity theft (\eg e-mail) \\

Terrorist attacks & Design against crime, VIVA, & \textbf{Anti-terrorist rubbish bins~\cite{doc_case_studies_2015}}: small bin mouth and volume prevents large disposals &  Logic bombs, watering hole \\
via station bins & CRAVED &  (\textbf{\emph{access}});  translucent, frequently monitored, and can be X-rayed (\textbf{\emph{visible/concealable}}); &  attacks \\
& & X-ray possible without triggering explosives, removable by a police robot (\textbf{\emph{value}}). & \\

Supply chain  & Situational crime prevention & \textbf{Multiple techniques~\cite{haelterman2013supply}}, \eg protect ground floor warehouse windows by anti-ram & Malware delivery, phishing,  \\
crimes & & posts (\textbf{increase effort}); screen consignments for prohibited articles (\textbf{increase risk}) & spam, session hijacking, \etc \\

Terrorism & Situational crime prevention & \textbf{Counter-terrorism SCP techniques and EVIL DONE~\cite{clarke2006outsmarting}} target prediction    & Cyberterrorism/hacktivism:   \\
& & framework (\textbf{E}xposed, \textbf{V}ital, \textbf{I}conic, \textbf{L}egitimate, \textbf{D}estructible, \textbf{O}ccupied, \textbf{N}ear, \textbf{E}asy) & DoS, malware, phishing, \etc \\

Identity theft  & Routine activity theory & Identity theft victimisation~\cite{reyns2013online}: 50\% more likely for online banking and e-mail/instant & Online identity theft \\
& & messaging users; 30\% more likely for online shopping and/or downloading behaviours & \\
& &  \textbf{$\Rightarrow$ focus on improving security for these services}. & \\

\bottomrule
\end{tabular}
}
\end{table}

\section{Adapting Environmental Criminology Concepts for Cyberspace}
\label{sec:principles_env_crim}

As we have seen, the environment is a crucial element to crime in the physical world.
For direct-contact crimes, it is imperative for a motivated offender and a victim or target to converge in space and time.
The environment influences the offender's decision whether to commit a crime or not, their modus operandi, and whether such crimes are likely to be repeated.
Therefore, altering the environment may be used to alter the decision-making of offenders, victims, or place managers, in order to prevent crime.
These principles have already been applied in the real world, but it is of great interest to see how they may be extended into cyberspace and for dealing with cybercrime more effectively.
In this section, we consider the key concepts of environmental criminology, and how they may be operationalised in the cyber realm.
Later, in Section~\ref{sec:cybercrime_place}, we focus on the concept of `place' in relation to cyberspace and propose how it could be adapted for cybercrime mitigation.

\subsection{Space and Time.}
\label{sec:space_time}
The spatial and temporal distributions of crime, which are based on the routine activities (or ``rhythms'') of law-abiding citizens and offenders alike, are the foundation of environmental criminological theory.
In this regard, the dimensions of space and time are critical in understanding why, where, and when crime occurs.
However, some aspects space and time differ significantly between the real world and the digital world, as do their effects on crime between these environments.

\noindent \textbf{The dimension of space.}
It is clear that the structure of cyberspace is considerably different to that of the real world. Whereas real space is continuous, the Internet is highly discretised with a node-edge topology (\eg traversing webpages through hyperlinks).
Cyberspace is also more ephemeral: websites can arise and disappear at rates much faster than land use in the real world.
Yar~\cite{yar_novelty_2005} particularly notes arguments that cyberspace universally has ``zero distance'' between its points, hence making it difficult to meaningfully translate physical concepts, such as proximity and location, to the analogous problem of crime in cyberspace.
Though there is indeed a theoretical basis for ``zero distance'' connectivity between computers, in reality, the concepts of proximity and location are still relevant in cyberspace for a number of reasons.
First, cyberspace has a firm rooting in the physical world.
The geographic locations of Internet service providers (ISPs), routers, and hosted web servers, and their relative connectivity, affect the structure of the Internet~\cite{tranos_death_2013}.
Political, economic, social, and cultural factors also affect the distribution of Internet infrastructure and usage.
Numerous examples exist, ranging from the differing amounts of Internet activity and connectivity across different social demographics and different regions~\cite{yar_novelty_2005}, to the intra- and international politico-economic factors that result in regional-specific restrictions on the Web (\eg nationwide censorship of the Internet).
Second, as Yar~\cite{yar_novelty_2005} notes, not all `places' are equidistant when negotiating the Web.
The ability of a user to find an entity on the Web greatly depends on how well other webpages reference that entity.
Therefore, destinations that require many hyperlink clicks and numerous hops could be considered relatively distant from a given starting point, in comparison to destinations that require fewer hops.
In this regard, the subjectivity of the user experience may be an important factor when considering distance in the cyber realm.

\begin{figure}[t]
    \centering
    \includegraphics[width=0.8\textwidth]{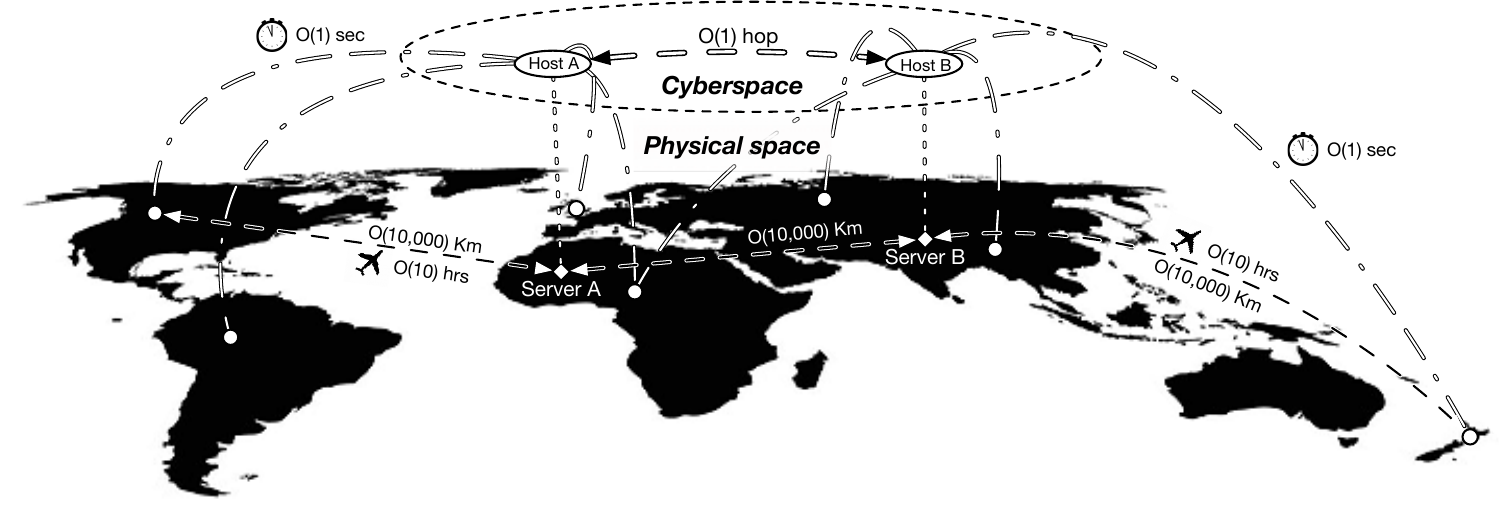}
    \caption{The contraction of distance and time in cyberspace.}
    \label{fig:contraction_space}
\end{figure}


One can also deduce indicators of space and distance in cyberspace from the activities and the interactions of Internet users, both offenders and non-offenders alike (which we do in Section~\ref{sec:cyber_enabled_dependent}).
Though the Internet allows for social networks to form irrespective of the physical distances between their members, one's physical proximities and relationships are likely to reflect in their online communications and activities~\cite{wellman_computer_2001,wellman_physical_2001}.
For example, one's e-mail, social network, or phone contacts, or the peers in their local area network, reflect their real-world relationships -- all of which may be exploited by, for instance, a motivated hacker or a cyberstalker.
As suggested by the geometric theory of crime~\cite{brantingham1993nodes,brantingham_criminality_1995}, offenders and victims are more likely to establish awareness spaces around online services that they frequent (irrespective of the associated risks) and, thus, are more comfortable in using~\cite{llinares_cybercrime_nodate}.
For example, users who visit pirate and infringing websites are less likely to install antivirus software, while, at the same time, being more exposed to malware~\cite{telang2018does}.

When discussing space and cyberspace, another important topic is that of the jurisdiction differential between states and countries, which is not as clearly apparent in the digital world.
That is, the laws and regulations of geographic regions enable some lines to be drawn on what constitutes acceptable behaviour and what criminal behaviour in the real world, as well as enabling the necessary agencies to enforce them.
However, with the ephemeral and cross-regional nature of cyberspace, drawing and enforcing these legal lines is an ongoing challenge~\cite{goodman2002emerging,llinares_cybercrime_nodate}.

It is worth noting at this point that little to no academic discourse regarding the juxtaposition between space and cyberspace (or `places' and `cyberplaces') and their relevance to cybercrime have ever been preceded by an exact definition of `place' in the first instance.
Perhaps, due to us being native to the physical world and not the digital world, we have taken the understanding of `place' for granted and deemed it too incongruent to be used in the cyber context without first addressing the ontological primitives of this concept, \eg the three components of physical places: \emph{location}, \emph{locale}, and the \emph{sense of place}~\cite{cresswell2014place}.
We elaborate on this in Section~\ref{sec:defining_cyberplace}.

\noindent \textbf{The dimension of time.}
What may first come to mind is the apparent instantaneity in the cause-and-effect of one's actions, such as the sending and receipt of an e-mail message, or the rapid execution of a complicated task by a computer program.
As Llinares and Johnson~\cite{llinares_cybercrime_nodate} (among others) note, this can make the subjective experience of time to appear shorter.
However, the temporal dimension is still important in determining Internet usage, as there is a direct relationship between the real-world activities of people (especially at a micro level) and their activities on the Internet.
For instance, users are more likely to access websites and download content for leisure and consumption outside of work hours~\cite{unpublished_ife_snapshot}.
Websites are more likely to be scheduled for maintenance late in the evening or during the early hours of the morning with respect to their local time zones.
Major real-world events are quickly followed by online news and social media chatter.

The time factor is also relevant to the interaction between offenders (or their actions) and victims.
For instance, users can only suffer threatening or abusive communications once they ``come online'' on a given service (\eg social media, chat messenger, forum), or once they access their e-mails.
Malvertisements and compromised webpages may remain dormant until a user accesses one of these pages before suffering a drive-by download attack.
Victims of phishing attacks only become so after opening the malicious e-mails.
Llinares and Johnson~\cite{llinares_cybercrime_nodate} note that, in general, peer-to-peer services can be characterised in one of two ways, based on the temporal mode of communication between Internet users.
The first type can be characterised as \emph{asynchronous} (store-and-forward or delayed) services, such as e-mail programs, mobile SMS, social networking sites and apps that facilitate direct messaging (Twitter, Facebook, Whatsapp \etc), and most static websites.
The second type can be characterised as \emph{synchronous} (real-time) services, such as VoIP services (Skype, FaceTime, Google Hangouts, \etc) and online multiplayer games that facilitate video, voice, and/or text-based chatting.
The time between the action of a offender and the consequent effect on a cybercrime victim ranges from an instant to several months, or even years.

\subsection{Offender Behaviours.}
\label{sec:offender_behaviours}
Following rational choice theory~\cite{clarke1985modeling} and its emphasis on offenders being rational decision-makers,
there are countless instances of cybercriminals and their malicious agents making rational selections of when, where, and upon whom they commit their crimes.
Intrusive cybercrimes are an example of the severe (IT-enabled) reduction in the costs and efforts associated with ``travelling'' and selecting suitable targets.
For example, when hackers or their malware agents infiltrate victim systems or networks, they often scan for vulnerable machines or sift through the contacts of victim accounts for further targets~\cite{hutchins2011intelligence}.
This ability results in a multiplying effect on the potential damage that these crimes may cause.
Furthermore, as IT enables users to easily connect with others on a global scale, cybercriminals may use this to their advantage in targeting as many victims as possible with the hopes of only successfully victimising a few in order to justify the effort. This approach is often the case with spam e-mail, botnet, and ransomware operations, which enable large-scale attacks with a few successful ones generating most of the revenue~\cite{stringhini_harvester_2014}.

In the real world, it has been shown that real or perceived anonymity may increase one's propensity to engage in antisocial behaviour in some circumstances (the Stanford Prison experiment~\cite{zimbardo1971stanford} comes to mind).
An interesting notion is the potential of anonymity in cyberspace to cause a similar increase in cybercriminal activity, though there is a present need for evidence-based studies to verify this.
Cybercriminals have also been shown to be aware of the risks of detection concerning their operations.
For instance, a large proportion of malware is capable of detecting antivirus software and honeypot environments (\ie dynamic malware analysis VMs) and, consequently, suspend their activities in order to hinder security analysis~\cite{chen2016advanced,bayer2009view}.
Malware also tends to utilise polymorphism (\ie the same malware family appearing in different guises) in order to avoid detection~\cite{bayer2009view}.
Some compromised or malicious websites conduct visitor fingerprinting to show specific pages to scrapers for search engine optimisation, malicious pages to end-users, and benign pages to potential security researchers and virtual machines~\cite{sood_crimeware-as--servicesurvey_2013}.
Cybercriminals tend to use evasive and anonymising techniques to prevent detection, such as by applying fast-flux~\cite{holz2008measuring} and domain generation algorithms (DGA)~\cite{antonakakis2012throw} techniques as part of their botnet infrastructures, hosting and conducting illegal business on Tor onion services, or using anonymous cryptocurrencies to carry out financial transactions (\eg collecting ransoms from compromised victims~\cite{kharraz2015cutting}), all in order to make it harder for law enforcement to catch those involved.

It is also interesting to consider how offenders become aware of criminal opportunities. As is the case in the real world, offenders are likely to form awareness spaces on the Web by way of the services that they regularly monitor or use.
This characteristic would allow offenders to become aware of criminal opportunities, such as bugs in a website or software, or the (vulnerable) demographic of users for a given service.
However, the Internet also affords offenders a higher level of surveillance to detect victims and criminal opportunity than in the real world, such as through the ability of a cyberstalker to observe when a user of a social networking app comes online, or a malware program to detect a working Internet connection on a victim's computer.
Hutchins~\etal~\cite{hutchins2011intelligence} describe this more formally as the `reconnaissance' stage of the cyber kill chain model.

\subsection{Suitable Targets.}
\label{sec:suitable_targets}
Environmental criminologists have assessed the applicability of the VIVA framework (Value, Inertia, Visibility, Access)~\cite{cohen_social_1979} to understanding how cybercriminals evaluate the suitability of potential targets~\cite{leukfeldt_applying_2016,yar_novelty_2005}.
In summary, they find that the dimensions of `value,' `visibility,' and `access' translate quite seamlessly from their physical interpretations to their digital ones.
For example, focusing on `value', a cybercriminal may assess the value of a target based on its financial potential, or its potential to increase the (notorious) reputation of the criminal.
Information security studies reveal a strong rational element in how cybercriminals assess their targets.
Paoli~\cite{de2018engineer} explicitly draws this out in his analysis of timesharing security engineers in the 1960s and 70s who, like criminologists, conceptualise malicious users
as rational actors who can assess the value of information.
Thomas~\etal~\cite{thomas2015framing} note that cybercriminals typically rent out compromised computers at varying prices depending on their regions, where computers from the West are usually more expensive in the cybercriminal economy than those from the rest of the world.
Concerning the use of stolen e-mail credentials, Onaolapo~\etal~\cite{onaolapo_what_2016} found that illicit users may ascertain the value of e-mail accounts by executing searches using keywords such as `bank' and `money.'
Turning to the other dimensions, the `visibility' of a target from the perspective of a cybercriminal could translate to a victim's online presence, or the presence of well-known vulnerabilities in a service (\eg a website bug, or a software CVE\footnote{\url{https://cve.mitre.org/}}).
The `accessibility' of a target may refer to a victim or system's attack surface by way of their software configuration (and associated vulnerabilities), or whether targeted data is stored within an access-controlled, digital environment or not.
Even the aspect of `inertia' -- the difficulty associated with an offender's ability to transport a physical good, or to overpower a victim, due to their mass -- is still relevant to cybercrime.
Though some have considered it ill-conditioned for cybercrime because of the apparent ``zero-mass'' of digital data, recent discussions~\cite{yar_novelty_2005,leukfeldt_applying_2016} have shown that the sizes of target data, and the (inhibited) technical specifications of a cybercriminal's computer, may be forms of inertia that can influence cybercrimes such as information theft.

\subsection{Guardianship and Natural Surveillance.}
\label{sec:guardianship}
Guardianship against crime is another relevant concept to the digital world.
Principally, family members, neighbours, or friends may act as protectors in the physical world for would-be victims of crimes such as cyber harassment and bullying~\cite{bossler2009line}. However, the concept of guardianship can be extended into cyberspace, both spatially (\ie guardians can operate over the Web), and in an anthropomorphic sense (\ie guardians may be software or `bots,' and not only human beings).
For instance, website owners, forum moderators, language filtering technologies can all act as guardians: to detect and prevent instances of cyber harassment and abuse.
Popular social media sites such as Facebook, Twitter, and YouTube automatically detect and remove explicit content.
Most Internet forums allow for the use of language filters to block inappropriate language.
Likewise, the concept of natural surveillance~\cite{newman1972crime} is also apparent in the cyber context, which talks of the ability of users to monitor spaces that they retain a shared interest.
Notably, within social networking, blogging, and e-commerce websites, ordinary users can report inappropriate, rule-infringing, and/or illegal posts and adverts to the moderators of these services.

For more technically advanced crimes, capable guardianship and place management
continue to be important to mitigating cybercrime.
Website and software users are empowered to report bugs for their remediation -- bugs which could otherwise be exploited and affect other users.
Bug bounty programmes~\cite{finifter2013bugbounty} create a financial incentive for such reports to be sent to the maintainer of the system and so allow the bug to be corrected.
Network administrators, security analysts, and their myriad of security technologies (firewalls, intrusion detection/prevention systems, \etc) are at the frontline of network-level protection, acting as guardians to users and devices within these networks.
Endpoint- and application-level guardianship is also present with operating system, antivirus, antimalware, and web application technologies such as spam filters, unsafe site alerts, and web application firewalls, all for the protection of the end-user.
In the mobile technology market, official app stores (\eg Apple Appstore, Google Play) are more likely to vet third-party apps for malware and employ stricter  development criteria than their unofficial counterparts~\cite{ng2014android}.
However, despite the proliferation of digital guardianship, the task of improving system security has shown itself to be a continuous arms race between security practitioners and cybercriminals.


\section{The Cybercrime Landscape and Information Security Mitigations}
\label{ref:infosec}
The field of environmental criminology has been, for the most part, primarily focused on crimes perpetrated in the physical world.
On the other hand, the information security community has been studying the different facets of cybercrime for decades.
Surprisingly, the parallel between the mitigations proposed by the information security communities and environmental criminology research was never made explicit.
The purposes of this section are threefold: (i) to give a general overview of the cybercrime landscape and current mitigations; (ii) to draw parallels between the mitigations proposed by the information security community and the theoretical models of environmental criminology; and, finally, (iii) to present some examples of new, potential mitigations by applying environmental criminology (Tables~\ref{table:new_mitigations} and~\ref{table:new_malware_mitigations}), which are at the end of this section.

\subsection{Anonymous Marketplaces}
With the ongoing rise in malware distribution, widespread data breaches, and the unethical collection and use of personal data by various corporations and governments, there has been widespread attention and development towards privacy-enhancing technologies and regulations.
One such technology that has become prominent is the Tor anonymous communication network.
This encrypted network is resistant to common Internet tracking methods and enables users (who utilise it correctly) to effectively remain anonymous from all but the most technically capable adversaries.
There are legitimate purposes for such a technology: users reading about sensitive topics, those with suppressed rights to freedom of expression,  journalism, whistleblowing, or those who object to targeted advertising.
Unfortunately, however, this anonymity has also been exploited to hide criminal activities, such as the trafficking of drugs, child sexual abuse images, violent pornography, and weapons.
Even worse, underground forums and anonymous marketplaces (\eg Silk Road) have arisen, enabling the convenient trade of such illicit products and services.
Researchers have also observed the rise of `crimeware-as-a-service' (CaaS) models~\cite{sood_crimeware-as--servicesurvey_2013} along with these ``underground markets''.
These criminal business models help to make cybercriminal operations (spam delivery, malware distribution, drug trafficking, money laundering) much more organised, automated, and accessible, especially for criminals with limited technical skills~\cite{stringhini_harvester_2014,sood_crimeware-as--servicesurvey_2013,christin2012traveling}.
Such business models have been made possible because cybercriminals can network with each other on these underground services and exploit various outsourcing opportunities.

\noindent \textbf{Mitigations:}
The primary methods of intervention towards illegal anonymous markets are server takedowns and arresting its operators.
These approaches were seen in law enforcement's takedown of the infamous Silk Road marketplace in 2013, which, at the time, was nearly a monopoly.
However, researchers have found that many more and diverse anonymous marketplaces have come to prominence since the takedown of Silk Road, with some (\eg Silk Road 2.0) arising in less than a month.
There is evidence of adaptation by these new marketplaces and their patrons, such as the increased use of encryption~\cite{soska2015measuring} and decentralised escrow services~\cite{horton2017hard}, and the diversification or specialisation in the types of products and services offered~\cite{dolliver2015evaluating,soska2015measuring}.
These changes mirror the well-known criminological mechanisms of \emph{crime displacement} (the net movement of crime elsewhere as a result of an intervention) and \emph{crime adaption} (cybercriminals altering their operations in order to bypass an intervention), which are potential, undesirable side effects of some interventions.

\subsection{Cryptocurrencies}
Decentralised cryptocurrencies have gained significant traction over the past decade, with Bitcoin being the first and most widely used cryptocurrency.
Bitcoin offers pseudonymity to its users, where accounts are not necessarily linked to real-world identities, but transaction details are publicly available in the distributed ledger.
Other cryptocurrencies, such as Zcash, are designed for full anonymity~\cite{bensasson2014zcash}.
Such properties are attractive to cybercriminals~\cite{brown2016cryptocurrency}, making cryptocurrencies popular for illegal activities, like purchasing illicit goods and services~\cite{meiklejohn2013fistful}, and enabling ransomware extortion~\cite{kharraz2015cutting}, digital theft~\cite{saito2015microeconomic}, and cryptocurrency laundering~\cite{bryans2014bitcoin}.
Kamps and Kleinberg~\cite{kamps2018moon} identified that cybercriminals take advantage of the unregulated nature of some cryptocurrencies to engage in ``pump-and-dump'' schemes.
This scheme is a type of fraud that involves three stages: accumulating a specific cryptocurrency coin, increasing its perceived value through misinformation (pumping), then selling it off to unsuspecting buyers at a premium price (dumping).

\noindent \textbf{Mitigations:}
Researchers such as Meiklejohn~\etal~\cite{meiklejohn2013fistful} and Harlev~\etal~\cite{harlev2018breaking} have devised techniques that can, to some extent, de-anonymise the operators of Bitcoin transactions.
Such techniques are especially useful for crime investigation and are similar to \emph{geographic profiling}~\cite{eck1995crime}, which involves connecting locations in a series of crimes by an offender in order to locate their ``anchor point'' (\eg their home).
These are also practical implementations of the \emph{`reducing anonymity'} situational crime prevention (SCP) technique, which increases the risks for cybercriminals by exposing their identities.
With regards to pump-and-dump schemes, Kamps and Kleinberg~\cite{kamps2018moon} devise an anomaly detection technique in order to identify these schemes within time-series data of the trading prices and volumes of different cryptocurrencies.
However, with an ever-increasing number of cryptocurrencies coming to the fore, and some that enable greater anonymity, it is clear that new approaches are needed to detect and discourage these sorts of criminal activities.

\subsection{Cyberbullying and Online Abuse}
With the advent of computer and networked technologies, the rapid adoption of the Internet has enhanced the abilities of end-users to perform their daily interactions -- communicating, purchasing and selling products, exchanging information, working, and engaging in leisurely activities -- without the limiting restrictions of time and space.
Likewise, there has also been an increase in criminal opportunity through such technologies, thus enabling and (potentially) multiplying crimes that traditionally relied on physical, human-to-human interaction.

Studies have followed the physical-digital transition of such interpersonal crimes and antisocial behaviour, like cyberbullying~\cite{navarro_going_2012,tokunaga_following_2010}, cyberstalking and cyberharassment~\cite{wick_patterns_2017,reyns_situational_2010}, online hate speech~\cite{mariconti_you_2018,hine2016kek}, and online child sexual exploitation and sexual harassment~\cite{henry_technology-facilitated_2018,acar_webcam_2017}.
These are only a few types of the crimes that have gained traction from such shifts in technology and society.

\noindent \textbf{Mitigations:}
The default mechanisms for dealing with online abuse (in its many forms) typically involve reporting abusive or offensive content (and their authors) to the relevant service moderators (or \emph{utilising place managers} from an SCP perspective).
In extreme cases, such as the commission of violent threats, online sexual harassment, or child sexual abuse images, one may report such behaviours to the police.
Although such actions can be useful, they are inherently reactive and often vulnerable to reporter biases (\eg opinions of inappropriacy, cultural differences) or false reporting, and are probably less effective in preventing future occurrences~\cite{mariconti_you_2018}.
Researchers such as Ioannou~\etal~\cite{andri_ioannou_risk_2018}, advocate the need for a proactive and multidisciplinary approach to dealing with online abuse.
Even automated filters, which ought to blacklist hate speech and offensive language, are limited, as in they rely on predefined dictionaries of words.
Such dictionaries are also inherently reactive and are inflexible towards misspellings and evolving language~\cite{serra_class-based_2017}.
Consequently, researchers have developed some proactive techniques for mitigating these crimes.

Mariconti~\etal~\cite{mariconti_you_2018} develop a supervised machine learning approach to automatically determine whether a YouTube video is likely to be ``raided'', \ie to receive sudden bursts of hateful comments.
Serra~\etal~\cite{serra_class-based_2017} propose a text classification algorithm using class-based prediction errors in order to more effectively detect evolving and misspelt hate speech.
Chatzakou~\etal~\cite{chatzakou_mean_2017} develop a system that automatically detects bullying and aggressive behaviour on Twitter, using text, user, and network-based attributes.
Founta~\etal~\cite{founta_unified_2018} present a holistic approach to automated abuse detection by supplying deep learning architectures with text and metadata-based inputs.
Yiallourou~\etal~\cite{yiallourou_detection_2017} devise a methodological approach that can be used to support the automated detection of images containing child-pornographic material.
The successes of such surveillance strengthening techniques, which are indeed subsets of risk-increasing SCP techniques,
are likely to increase the risk of getting caught for offenders and are just some of the multidisciplinary ways to deal with such problems.
Of course, other forms of countermeasures exist.
For example, the impersonation of minors by law enforcement has been shown to be effective in apprehending offenders, while automated chatbots are being developed to profile potential offenders~\cite{acar_webcam_2017}.
There is also the arrest and prosecution of the worst offenders~\cite{endrass2009consumption}.
Educating minors and Internet users to avoid online abuse victimisation is also an important, long-term initiative~\cite{wolak2008online}.

\subsection{Cyber Fraud}
Fiancial crime and fraud have also made a paradigm shift into the cyber world.
The phenomenon of advance-fee fraud, or ``419'' scams (cybercriminals reaching out to potential victims with grandiose promises of wealth in exchange for advanced payments from them) have been well-documented by researchers~\cite{herley_why_2012}.
Recent works have found such scams are more of a universal issue than once thought~\cite{mba_flipping_2017}, rather than being one that only involves less economically developed countries.
Cybercriminals have also been known to target other services for fraudulent activities, depending on their demographics of interest.
For example, ``419'' scams are likely to be delivered en masse through spam e-mail communications, where gullible recipients would self-identify themselves by responding to these e-mails~\cite{herley_why_2012}.
Romance scammers are likely to operate on dating websites in order to manipulate emotionally vulnerable users~\cite{edwards_geography_2018,huang2015quit,buchanan_online_2014}.
Consumer fraudsters are likely to target large online marketplaces to commit buyer or seller fraud~\cite{van2011bought}.
Various forms of identity fraud, facilitated through Internet-enabled theft of personally identifiably information (PII) (\eg names, addresses, e-mail addresses) or account credentials for common services (\eg e-mail, banking, social media) are also problems that the information security community closely monitor.
Researchers have recognised that phishing e-mails and malware are common precursors to identity fraud~\cite{roberts_fear_2013}, and they have monitored the illegal activities that subsequently ensue with such credentials~\cite{onaolapo_what_2016}.

\noindent \textbf{Mitigations:}
The effective prosecution of scammers is necessary but often difficult due to the transnational nature of these operations and the relatively small amounts of money involved per fraud.
Some engineering countermeasures are in use, such as the use of spam or phishing filters to prevent malicious messages reaching recipients, or blacklists that raise alerts or block known phishing websites.
However, the maintenance of such measures is a continual arms race, as cybercriminals are always adapting these spam messages or compromising new websites to avoid these blockers.
It is possible for services to automatically detect scammer profiles, such as by their reuse of profile descriptions or profile photos~\cite{edwards_geography_2018}.
On the other hand, perpetrators could also adapt to such countermeasures with ease.
Arguably, the most effective countermeasures could be to reduce the profitability, or increase the required effort, for such crimes.
An economic strategy, such as increasing the transaction fees or the necessary background checks for money transfer services, could be a set of mitigations that attack the profitability of such crimes.
Awareness campaigns could also help to reduce the opportunity for victimisation, but perhaps more so if these campaigns are directed towards the most vulnerable, as identified by their personality types and victimisation statistics~\cite{buchanan_online_2014,whitty_online_2012}.
With regards to environmental criminology, these are recognised as \emph{market disrupting} and \emph{target removing} SCP techniques, which involve reducing the rewards of crime by denying criminals the ability to steal, sell, or access a target.




\subsection{Malware and Botnet Operations}
One area of focus in the information security community is the study of malicious software -- malware.
The issue of malware came into prominence in the 1980s, but in recent times, it has become a massive underground economy.
In short, financial motivations (above others) have become a cornerstone to the design and proliferation of modern malware.
Researchers have identified that modern strains typically carry a myriad of functions, no doubt for the purposes of monetisation.
Malware families, such as Zeus~\cite{binsalleeh_analysis_2010}, for example, can steal banking and financial credentials on compromised machines, log keystrokes and extract documents, or to encrypt victim computers to be held for ransom.
Even worse, some malware families are designed to retain control of compromised devices in order to assimilate them into larger networks of infected machines, or botnets. These botnets may be used (or rented as-a-service) to facilitate distributed denial-of-service (DDoS) attacks against a target, to send spam e-mails~\cite{stringhini_harvester_2014}, or to mine cryptocurrencies.

Malware distribution has been refined to infect as many viable victims as possible.
Initially, there was a heavy reliance on human activity and manipulation, such as the need for victims to open spam e-mail messages~\cite{stone2011analysis} or to be social engineered into activating a malicious file~\cite{nelms2016towards}.
Nowadays, cybercriminals have developed distribution mechanisms to completely cut out the need for human interaction, such as delivering malware directly through automated browser-based attacks (or drive-by download attacks) via compromised websites or malvertisements~\cite{zarras2014dark,nappa_driving_2013,sood_malvertising_2011}.
To ease the lives of malware operators, the cybercrime ecosystem proceeded to come up with exploit kits -- software packages that deliver a wide variety of exploits for different computer configurations~\cite{grier2012manufacturing}.
This innovation, ultimately, increases the probability of a victim's system becoming compromised.
In a further attempt to streamline malware delivery and lower the entry bar for cybercriminals, pay-per-install (PPI) schemes have also arisen in the cybercrime ecosystem~\cite{caballero2011measuring}.
These services are specialised botnets of infected devices that enable the distribution and download of new malware onto these already compromised machines.
PPIs are set up and managed by a service provider, whom customers pay in order to infect machines with their own proprietary malware.

The disruption of the malware distribution economy is an ongoing challenge.
Cybercriminals increasingly implement new and numerous techniques in order to prevent their malware and botnets from being detected and disabled.
Researchers have found that malware often obfuscates their outgoing communications, undergo polymorphism to ``change their appearances'', remain ``silent'' whenever they detect a possible malware analysis environment, copy themselves to multiple locations on a compromised machine, or distribute themselves over multiple devices on a network~\cite{bayer2009view}.
Botnet operators have also been found to employ various tactics to avoid detection and takedown attempts of their infrastructures, such as implementing fast-flux techniques (the rapid rotation of IP addresses)~\cite{holz2008measuring}, or domain generation algorithm techniques (the constant changing of domain names)~\cite{antonakakis2012throw}, to hide the locations of their command and control servers.

\noindent \textbf{Mitigations:}
The challenges of malware and botnet infrastructures are as complex as their operations.
First, there is the issue of preventing malware infection and spread.
Signature-based antivirus programs have long been the major defence in detecting and removing malware, along with intrusion detection systems and content filters.
However, they struggle with the extensive manner of forms that malware now appears (polymorphic, metamorphic, compressed, encrypted, \etc).
Antivirus programs that use heuristic methods for malware removal are now much more common (\ie detection is based on abnormal program behaviours).
Notwithstanding, malware removal is still a reactive strategy, so proactive measures have also been developed.
One such is the use of antimalware tools, which attempt to prevent malware attacks in the first instance through methods such as malware sandboxing, raising browser alerts on suspicious websites, and preventing the spread of malware if a device is infected.
Another proactive approach is vulnerability assessment and management, which deals with providing regular system updates in order to remove known vulnerabilities.
Such updates would reduce the success of drive-by download attacks, for example, thus minimising one's attack surface for malicious actors to exploit.
Both of these techniques are akin to \emph{target hardening} that is applied in SCP and CPTE/UD frameworks, which aim to increase the difficulty of an attacker gaining access to their target.
Although malware delivery is not completely dependent on human error, this role is still substantial.
Educating users to keep their systems up-to-date and on how to avoid social engineering attacks are some non-technical approaches that are also applied, such as with Action Fraud and their \emph{\#UrbanFraudMyths}\footnote{https://twitter.com/hashtag/urbanfraudmyths}.

Second, there is the issue of disrupting botnet operations.
One important technique involves the infiltration of botnets by security researchers~\cite{binsalleeh_analysis_2010,abu_rajab_multifaceted_2006,cho2010insights,caballero2009dispatcher}.
Such techniques allow researchers to collect intelligence on cybercriminal operations, and identify weak points in their communication protocols for disruption, or locating their C\&C servers for ISP takedowns.
They may also be used to identify the owners of these botnets, such that law enforcement may arrest and prosecute them.
However, with the estimates of Kaspersky Lab~\cite{noauthor_botnet_2018} indicating there could be hundreds of thousands of botnets in the wild, it is difficult to see the scalability of these techniques.
Alternatively, service providers may provide some mitigations.
For example, e-mail programs and social networking sites usually employ spam filters, which may consequently deter spam operations.
However, these filters are often signature-based, so minor adjustments in the spam messages may cause them to go undetected.
ISPs may use DNS sink-holing techniques and blacklists to prevent their customers from accessing sites known to be malicious.
However, such techniques also come under the ``arms race'' issue of keeping up with the cybercriminals.
Other economic measures are possible, such as pressuring ISP services to dissociate from ``bulletproof ISPs'', which resist law enforcement and typically harbour these criminal activities, or pressuring financial institutions to dissociate from rogue banks, which liaise with cybercriminals, in order to effectively shut down their operations.
Environmental criminology recognises these as \emph{market disrupting} techniques (SCP), which aim to reduce the economic benefits of such operations until they are no longer viable.

Using the SCP framework, a proof-of-concept matrix of potential mitigations for disrupting malware operations is provided in Table~\ref{table:new_malware_mitigations}.

\subsection{A Synergistic Approach}
Though there are already clear parallels between the theoretical models of environmental criminology and the mitigative techniques proposed by information security, security researchers are yet to fully explore the \emph{structured} analytical and actionable processes that environmental criminology has to offer.
Firstly, past and current mitigations devised by security researchers only seem to represent or consider a subset of the full range of techniques that could be utilised, while lacking a systematic approach to establish these.
Secondly, little attention seems to be directed towards the consideration, monitoring, and evaluation of the actual effects of interventions by security researchers, both with regards to the victims/targets and the malicious actors, and how they respond.
Ultimately, without considering the fulness of the crime prevention process, mitigations are more likely to fail (to different degrees) in the goal of controlling crime in both the short- and long-term, as cybercriminals may quickly identify alternative targets, crime types, or modus operandi.

\begin{table}
\centering
\caption{Some examples of cybercrime countermeasures using environmental criminology.}
\label{table:new_mitigations}
\resizebox{1\linewidth}{!}{
\begin{tabular}{l|l|l}
\toprule
\textbf{Cybercrime} & \textbf{Technique} & \textbf{New Countermeasure} \\
\midrule \midrule

Darkweb market  & Situational crime prevention & \textit{Increase the perceived risk} of incarceration for Darkweb market users by increasing the \\
solicitation &  & perceived number of undercover law enforcement agents within them and the awareness \\
&  &  of their presence. Also, raise publicity of operations targeting low-level cybercriminals, \\
& &  as well as high-level ones. \\
& & \textit{Increase the risks/Reduce the rewards:} introduce significant numbers of honeypot credit cards \\
& & and email accounts under law enforcement control into underground circulation \\
& & (dark forums/markets, surface web) \\
\midrule


Cyberbullying  & Crime prevention through & \textit{Peer-to-peer monitoring}: users are periodically asked to review the anonymised messages  \\
and online abuse & environmental design & (\ie PII redacted) involving other users, who are either random users or members of the\\
&  &    recipient's network, and confirming if these messages are rule-infringing or not (and why).\\
&  &  This system could help to identify malicious communications and protect vulnerable users. \\
\midrule

Romance scams  & Crime scripting / & STEP 1: fraudsters create a fake account $\Rightarrow$ \textit{reduce anonymity:} require ID for registration; \\
 & Situational crime prevention & \textit{utilise place managers:} identify and flag potential ``predator'' profiles \\
 &  & STEP 2: identify a suitable victim $\Rightarrow$ \textit{conceal targets:} identify vulnerable users (\eg survey  \\
& & on registration) and offer increased protection (\eg reduced visibility to unscrutinised users); \\
& & STEP 3: groom victims $\Rightarrow$ \textit{target harden:} educate users to identify common romance scam \\
& & modus operandi (periodic awareness reminders and tests for users to identify scammers) \\
& & STEP 4: commit fraud by asking victim for money $\Rightarrow$ \textit{target harden} as above. \\


\bottomrule
\end{tabular}
}
\end{table}

\begin{landscape}

\begin{table}[p]
\centering
\resizebox{0.9\linewidth}{!}{
\begin{tabular}{p{5cm}||p{5cm}|p{5cm}|p{5cm}|p{5cm}}
\toprule

\footnotesize \textbf{Malware Value Chain} 
& \footnotesize \textbf{\textit{Increase the perceived effort}} 
& \footnotesize \textbf{\textit{Increase the perceived risks}} 
& \footnotesize \textbf{\textit{Reduce the anticipated rewards}} 
& \footnotesize \textbf{\textit{Remove the excuses for crime}} \\

\midrule \midrule


\multirow{5}{5cm}{\footnotesize\textbf{\textit{1. Malware development}} \\ 
\footnotesize - malicious code-sharing \\ 
\footnotesize - outsourcing malware development \\ 
\footnotesize - engaging crimeware-as-a-service (CaaS) providers
}
& \multirow{2}{5cm}{\footnotesize - Disrupt dark markets and cybercriminal communication channels. \\
\footnotesize - Infiltrate dark markets.
}
& \multirow{3}{5cm}{\footnotesize - Publicise reverse-engineered malware and remediation. \\
\footnotesize - Publicise LEA operations in dark markets. \\
\footnotesize - Upscale malware analysis honeypots.
}
& \footnotesize - Publicise reverse-engineered malware and resulting remediation.
& \multirow{2}{5cm}{\footnotesize - Publicise / educate public on the impacts of malware on victims. \\
\footnotesize - Increase reverse-engineering job roles. \\ 
\footnotesize - ``Malware bounty hunting`` programmes.
} \\

& & & & \\ & & & & \\ & & & & \\

\midrule

\multirow{12}{5cm}{\footnotesize \textbf{\textit{2. Malware delivery}} \\
\footnotesize - Manual infections \\
\footnotesize (social engineering, physical access) \\
\footnotesize - Automated infections \\
\footnotesize (exploit kits, mass emails, network diffusion) \\
\footnotesize - Delivery networks \\
\footnotesize (pay-per-install (PPI) networks, CaaS botnets)
}
& \multirow{12}{5cm}{\footnotesize - Educate public on cybersecurity. \\
\footnotesize - Harden targets (application, host, network, and Internet sec.). \\
\footnotesize - Improve ISP security practices (websites, CDNs, cloud hosting, broadband). \\
\footnotesize - Regulate ISPs and standardise security protocols. \\
\footnotesize - Regulate and police software advertisers and resellers (PPIs). \\
\footnotesize - Infiltrate PPIs and botnets. \\
\footnotesize - Takedown botnets (arrest, seizure, sinkhole). \\
\footnotesize - Pressurise ``bulletproof'' ISPs (C\&Cs).
}
& \multirow{12}{5cm}{\footnotesize - Malware / exploit kit detection. \\
\footnotesize - Increase monitoring of CDNs and software marketplaces (\eg GitHub, App Stores). \\
\footnotesize - Publish security advisories for malicious software. \\
\footnotesize - Less distinguishable malware honeypots. \\
\footnotesize - ``Malware bounty hunting'' programmes. \\
\footnotesize - Regulate and police software advertisers and resellers (PPIs). \\
\footnotesize - Infiltrate PPIs and botnets. \\
\footnotesize - Takedown botnets (arrest, seizure, sinkhole).
}
& \multirow{12}{5cm}{\footnotesize - Educate public on cybersecurity. \\
\footnotesize - Virtualisation / ephemeral sessions. \\
\footnotesize - Harden targets (application, host, network, and Internet sec.). \\
\footnotesize - Less distinguishable malware honeypots. \\
\footnotesize - Malware ``vaccination'' and disinfection campaigns. \\
\footnotesize - Regulate and police software advertisement and reselling industry (PPIs). \\
\footnotesize - Takedown botnets (arrest, seizure, sinkhole). 
}
& \multirow{6}{5cm}{\footnotesize - Vulnerability bounty programmes (hosts, websites, web applications). \\
\footnotesize - ``Malware bounty hunting'' programmes. \\
\footnotesize - Publicise / educate public on the impacts of malware on victims.
}\\

& & & & \\ & & & & \\ & & & & \\ & & & & \\ & & & & \\ & & & & \\ & & & & \\ & & & & \\ & & & & \\ & & & & \\ & & & & \\ & & & & \\

\midrule

\multirow{11}{5cm}{\footnotesize \textbf{\textit{3. Post-infection activities}} \\
\footnotesize - Botnet control, CaaS \\
\footnotesize - Secondary cybercrimes \\
\footnotesize (data theft, financial fraud, ransomware, replication, spam emails, crypto-mining, DDoS, PPI)
}
& \multirow{9}{5cm}{\footnotesize - Infiltrate botnets/CaaS services. \\
\footnotesize - Takedown botnets/CaaS services. \\
\footnotesize - Improve key service provider security and fraud detection (online banking, e-commerce, cloud services). \\
\footnotesize - Regulate key service providers and standardise security protocols. \\
\footnotesize - Implement DDoS prevention at ISP-level (broadband). \\
\footnotesize - Pressurise ``bulletproof'' ISPs (C\&Cs).
}
& \multirow{5}{5cm}{\footnotesize - Infiltrate botnets/CaaS services. \\
\footnotesize - Takedown botnets/CaaS services. \\
\footnotesize - Monitor aggregate network/activity activity (service providers) for anomalies.
}
&\multirow{11}{5cm}{\footnotesize - Takedown botnets/CaaS services. \\
\footnotesize - Malware ``vaccination'' and disinfection campaigns. \\
\footnotesize - Virtualisation / ephemeral sessions. \\
\footnotesize - Harden targets (data encryption). \\
\footnotesize - Automated off-site backups. \\
\footnotesize - Educate public on cybersecurity. \\
\footnotesize - Improve key service provider security and fraud detection. \\
\footnotesize - Regulate key service providers and standardise security protocols. \\
\footnotesize - Implement DDoS prevention at ISP-level.
}
& \multirow{1}{5cm}{\footnotesize - Publicise / educate public on the impacts of malware on victims.
}\\

& & & & \\ & & & & \\ & & & & \\ & & & & \\ & & & & \\ & & & & \\ & & & & \\ & & & & \\ & & & & \\ & & & & \\ & & & & \\

\midrule

\multirow{4}{5cm}{\footnotesize \textbf{\textit{4. Post-malware activities}} \\
\footnotesize - Consumption of stolen data \\
\footnotesize - Beneficiaries of damage \\
\footnotesize - Monetisation and laundering
}
& \multirow{4}{5cm}{\footnotesize - Disrupt dark markets. \\
\footnotesize - Regulate financial/crypto services. \\
\footnotesize - Pressurise shady financial services.
}
& -
& \multirow{4}{5cm}{\footnotesize - Regulate financial/crypto services. \\
\footnotesize - Pressurise shady financial services.
}
& - \\

& & & & \\ & & & & \\ & & & & \\


\bottomrule
\end{tabular}
}
\caption{A high-level malware value chain (left column) and a matrix of potential countermeasures using Situational Crime Prevention. \\
N.B: the countermeasures proposed are not exhaustive -- \eg the \textit{`reduce the provocations'} category was omitted.}
\label{table:new_malware_mitigations}
\end{table}

\end{landscape}

\section{Adapting the Concept of Place for Cyberspace}
\label{sec:cybercrime_place}

It is apparent that the environmental aspects and criminological principles that exist with physical crime can, to a large extent, be extended and applied to cybercrime.
However, there is still the need for a clear conceptualisation of `place' with respect to the digital space and cybercrime, since the concept of place is key to environmental criminology.
In this section, we deduce what `place' means in the context of cybercrime by examining various types of cybercrimes, and also by considering `place' in the real world.
We use the classifications of \emph{cyber-enabled} crimes and \emph{cyber-dependent} crimes to assess the concept of `place' within each class of cybercrime, as defined by criminologists~\cite{mcguire2013cyber,leukfeldt_applying_2016}.

\subsection{Analysing Cyber-Enabled and Cyber-Dependent Crime Contexts}
\label{sec:cyber_enabled_dependent}
Cyber-enabled crimes~\cite{mcguire2013cyber} are crimes that occur in the real world but can be enhanced, expanded, and optimised by Internet technologies.
That is, the Internet makes it is easier and cheaper for cybercriminals to find victims, to operate internationally, and to avoid getting caught.
Some cyber-enabled crimes include identity theft, consumer fraud, various forms of cyber harassment and threatening communications, and the trafficking of illegal products or services through the Internet.
Cyber-dependent crimes~\cite{mcguire2013cyber}, on the other hand, refer to crimes that are only possible as a result of computer and networked technologies, such as hacking, malware infection, or botnet operations.
Though these crimes will still have real-world consequences (\eg identity theft, financial fraud), they can only occur through computers.
Just as physical crimes occur in particular places, we are able to review a wide variety of cybercrimes and identify their associated `cyberplaces,' especially focusing on the crime event and the context in which it is commissioned.

\subsubsection{Cyber-Enabled Crimes.}
Researchers have found that interpersonal crimes (cyber harassment, cyberbullying, cyberstalking, online sexual exploitation, \etc) primarily occur on online services ranging from e-mail, mobile phone, and instant messengers~\cite{wick_patterns_2017,tokunaga_following_2010} to blog sites, social networking sites, forums, and chat rooms~\cite{reyns_situational_2010}, to online games~\cite{henry_technology-facilitated_2018} and various VoIP technologies (\ie real-time video chat)~\cite{acar_webcam_2017}.
Financially-motivated cybercrimes have been found to occur also on these same online services, though the purposes and modus operandi of these crimes differ to interpersonal crimes.
Whereas interpersonal crimes primarily involve the use of these services to emotionally traumatise and instil fear in victims, financial crimes, such as identity theft or consumer fraud, primarily involve the use of these services to induce potential victims for information and financial theft.
For example, phishing scams~\cite{lynch_identity_2005} and advance-fee ``419'' scams~\cite{mba_flipping_2017,herley_why_2012} typically occur on e-mail services and web forums.
Other scams that are unique to a type of online service, such as romance scams~\cite{edwards_geography_2018,whitty_online_2012} or buyer or seller fraud~\cite{van2011bought}, almost always begin with the cybercriminal contacting the victim through these services (match.com, eBay, \etc) and luring them into offline communications (e-mail, telephone, or in-person), before defrauding their victims or committing other crimes.
Profile name re-users, or ``impersonators'' -- people who take advantage of reputable profile names because of their large followings -- operate on social networking sites in order to engage their newly acquired followers in illegal activity (spam, illegal sales)~\cite{mariconti_whats_2017}.

Further still, crimes that involve the trafficking of illegal products and/or services (drugs, weapons, child sexual abuse images, malware, stolen credentials and credit cards, \etc) also occur on online services, including through anonymised networks such as Tor, and using pseudonymous cryptocurrencies such as Bitcoin~\cite{celestini2017tor,christin2012traveling} for conducting transactions.
These services allow perpetrators to take advantage of the transnational nature of the Internet to upscale their distribution networks, their consumer markets, and, ultimately, their financial profits.
Here, it is clear that the concept of `place' is inherent: people know the sites to visit and the applications to use in order to carry out their online activities, whether legal or not.

\subsubsection{Cyber-Dependent Crimes.}
The nature of cyber-dependent (or ``high-tech'') crimes differ to that of interpersonal crimes.
Primarily, cyber-dependent crimes exist only due to the presence of computers and the Internet, whereas interpersonal crimes (harassment, consumer fraud, \etc) already exist in the real world, but are only amplified through such technologies.
More precisely, the primary target of high-tech crimes (hacking, malware, denial-of-service attacks, \etc) are the computers and digital resources themselves, before the physical users who own or use them.
Because of this, such crimes can, at times, bypass the need for human activity in order to occur.
For example, security researchers have identified buffer overflow attacks (which occur on victim computers) as a common attack vector used by malware, such as the \emph{Blaster} and \emph{Code-Red} worm families~\cite{bailey_blaster_2005,moore_code-red:_2002}, completely bypassing the need for social engineering.
Network protocol flaws have been exploited in order to establish malicious network connections, break into systems, and automatically spread malware throughout a victim's local network~\cite{koniaris_honeypots_2014,morales_analyzing_2010}.

Nonetheless, the ``human element'' is still exploited whenever possible.
A significant proportion of malware is delivered to victim computers through social engineering, such as in the download of malicious e-mail attachments.
For example, the \emph{Bagle.AH} and \emph{Netsky.C} worms propagate themselves as e-mail attachments to addresses that they find on infected computers, while also spreading through file-sharing peer-to-peer networks and local network drives~\cite{bose_agent-based_2011}.
Victims are also directed to malicious websites, via e-mail, social media~\cite{mariconti_whats_2017}, or through malvertisements~\cite{sood_malvertising_2011}, where they consequently face bombardment by silent drive-by download attacks.
Such attacks often result in the forced download of malware onto the victims' computers~\cite{nappa_driving_2013,sood_crimeware-as--servicesurvey_2013}.
The websites themselves that utilise vulnerable software are also targeted, such as through SQL injection and XSS attacks.
In particular, websites that use content management systems, such as Wordpress, are more likely to be targeted and compromised than others due to their higher market share~\cite{vasek_hacking_2016,vasek_identifying_2014}.
Mobile users are not exempt from such devious strategies. For instance, security researchers have found that the \emph{Mabir} worm spreads through bluetooth and MMS, prompting nearby potential victims to accept its installation~\cite{bose_agent-based_2011}, while the \emph{Geinimi} trojan app is installed through third-party app stores, which consequently opens back doors on these devices and exfiltrates information~\cite{choo_cyber_2011}.

There is also a relationship between the virtual places and geographic locations where cybercriminals tend to carry out their nefarious activities.
For instance, the assimilation of compromised machines into botnets is common after malware exploitation.
The prices of these `bots' (hence their values) vary in underground markets based on the countries where they are located~\cite{thomas2015framing}.
The command and control servers that are used to monitor and operate these botnets are typically hosted by ``bulletproof'' ISPs, which are based in countries resistant to law enforcement pressure and take-down attempts~\cite{sood_crimeware-as--servicesurvey_2013}.
Similarly, malware that spreads through a host tends to look for peers that share some proximity to the victim, \eg via e-mail address books, shared computer networks, or social media connections.
More generally, every computer and network resource is tied to physical infrastructure (\eg computer memory, physical devices) that is located in the real-world, each presenting unique opportunities for crime.

In the context of these crimes, `place' is more broad in scope, from websites and webpages to online services and web applications, to computer devices and their software, peripherals, and networks.
These examples exhibit a marked difference in how `place' is perceived between cyber-enabled and cyber-dependent crimes, where the former gravitate around online services and web applications (human-to-human activity), while the latter permeates every area of computing.

\subsection{A Framework for Defining Cyberplace}
\label{sec:defining_cyberplace}
The compatibility of environmental criminology with the practicalities of cyberspace merit, we believe, a new, complementary research direction towards mitigating cybercrime.
Moving forward, the key to this new research direction is the development of a consistent definition of `place' in cyberspace, or, simply, \emph{cyberplace}.
We have shown that the concept of place in cyberspace is strongly apparent on both the user- and device-levels: Internet users know the websites to visit and the applications to utilise for their work, leisure, or consumptive activities.
Cybercriminals know the services to use to find and exploit their victims, to target assets, or to engage and trade with other cybercriminals.
Routing devices know how and where to transmit information, through various network protocols (\eg TCP/IP) in order to reach any part of the world.
Devices within a computer network are communicable by their own internal IP and MAC addresses.
Even internally, every computer application, library, instruction, and chunk of data is addressed in memory and can, thus, be pinpointed to a physical device in the real world.

We have also shown that the way that each cyberplace is considered can be contextual, depending on the perspective of the actors who interact with them.
For example, a web domain hosting a chatroom may constitute a single cyberplace for its users or a crawler bot, but in a networked sense (\eg a router or a DNS resolver), it could represent different cyberplaces if it undergoes changes in its public IP address.
Characterising these different cyberplaces would help the adaptation of crime prevention frameworks towards analysing and mitigating cybercrime more effectively.

\noindent \textbf{Revisiting the real-world.}
Though we have spent considerable effort in identifying different cyberplaces through the contexts of cyber-enabled and cyber-dependent crimes, it is also worthwhile revisiting the primitive concept of `place' from the real-world perspective.
`Place' has long been used and commonly understood within society for millenia, but it is only in the last few decades that geographers have conceptualised it as a particular location that has acquired a set of meanings and attachments~\cite{cresswell2014place}.
It is recognised that place can be conceptualised in terms of the social interactions that they tie together, in that people go to places to engage in activities and that they interact with places themselves~\cite{massey2013space,cresswell2014place}.
However, more concretely, `place' can be considered as a meaningful site that combines three fundamental components: \emph{location}, \emph{locale}, and \emph{a sense of place}~\cite{cresswell2014place}.
Location refers to the ``where'' or the absolute position of a place (\eg the geographical coordinates of a university library).
Locale refers to its material and tangible setting -- the way that a place looks and what is contained therein (\eg the university library on a busy street in central London, surrounded by buildings, having a gate and a courtyard, and long corridors and thousands of books within it).
Finally, the sense of place refers to the abstract feelings and emotions associated with that place, which may be derived individually or by shared experiences (\eg it is a place for study, there is a presumed abundance of written materials, and there are personal experiences and history associated with the library).
With these three elements, people (unconsciously) identify and differentiate places in the physical world.
In a similar way,
we propose that cyberplaces may be conceptualised as a combination of three fundamental components: \emph{location}, \emph{state}, and \emph{function} (Figure~\ref{fig:cyberplace}), which we describe as follows:

\begin{figure}[t]
    \centering
    \includegraphics[width=0.5\textwidth]{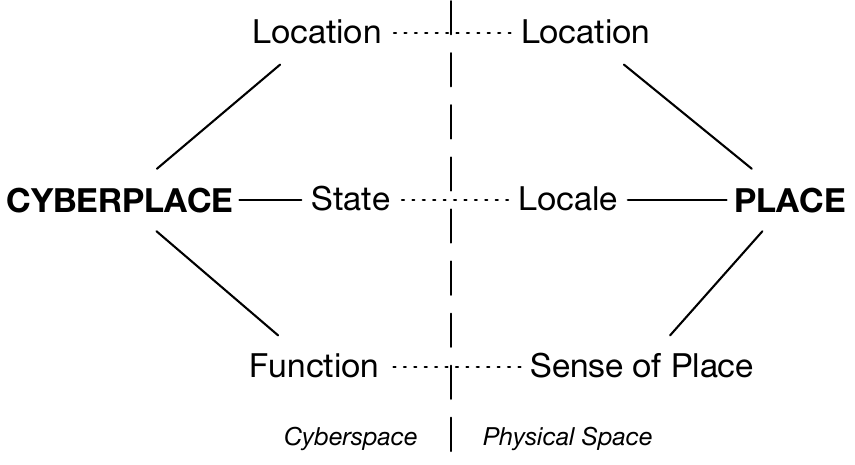}
    \caption{The analogous concepts of `cyberplace' and `place.'}
    \label{fig:cyberplace}
\end{figure}

\subsubsection{The \textbf{Locational} Component.}
This component is analogous to \emph{location} with respect to physical places~\cite{cresswell2014place}, which refers to the precise point or GPS location of a physical place.
Thus, particularly in relation to how information flows within and between computers, the locational component of the cyberplace concept encapsulates information pertaining to the specific locations of cyberplaces in terms of digital address spaces.
These address spaces encompass entire networks and computer hosts (\eg IP and MAC addresses), peripheral devices, disk sectors, and even the memory cells within a computer.
Therefore, just as computers and networks can inherently resolve the URL of a page on a website to a specific IP address and file directory, or resolve a function call in a software program to a specific address  within a computer's address space, locational information enables us to specify the precise location of each cyberplace in cyberspace.
In the same way, a change in ``location'' (address) will constitute a change in the overall cyberplace, which may be perceptible, such as would be in the case of a server changing its IP address, where the relevant DNS servers (the perceiving actors) would update their address mapping (A) records for this domain and new IP.
However, such changes may not be perceptible to actors who just access a server through its URL.

The locational component also explicitly links each cyberplace to a real-world location and device.
For example, every website has an IP address, which can be traced to a physical server.
Every instance of a file or an application can be traced to one or more memory locations within a computer system or peripheral device.
Thus, this component provides useful information for the stage of implementing real-world interventions, such as pinpointing the real-world locations of cybercriminals and their servers for arrest and takedown.
Of course, it is worth noting that the issue of criminals masquerading (or spoofing) their device locations is an operational issue, based on Internet design, and not a conceptual one.
As we have previously discussed, `location' can be an important factor from the criminal prospective (\eg cybercriminals seeking targets from a certain country) and it is recognised by both human and machine actors (computers, malware, bots).


\subsubsection{The \textbf{Statal} Component.}
The second component of the cyberplace concept is the statal component, which encapsulates information relating to the state of a cyberplace and its tangible aspects at a specific point in time.
This is analogous to the \emph{locale} aspect of physical places~\cite{cresswell2014place}, which refers to the way a place looks, its tangible aspects, and the surrounding environment.
The tangible aspects of a cyberplace can be wide-ranging, and both internal and external to its environment.
The internal state of a cyberplace (and how it is perceived from within) could be attributable to, for example, the content and hyperlinks on a webpage, or the webpage or software source code and the external resources that it utilises, or the files within a directory, or the overall design and appearance of the cyberplace.
On the other hand, the extenal state of a cyberplace (and its visibility) could be assessed by, for example, the distribution of hyperlinks from external sites leading to a particular webpage (which affects the perceived distance from these sites to this cyberplace), or the popularity and generated traffic of a webpage (which may be captured with Alexa\footnote{\url{https://www.alexa.com/siteinfo}} or search engine rankings), or, in the case of a computer system, an application, or a process, whether a virtualised environment or the presence of antivirus, firewalls, or intrusion detection systems can be detected (\eg a malware deciding whether to operate or hide its functionality).

The statal component is important because human and machine actors alike have some ability to assess the immediate context of a cyberplace (\eg a webpage, a web application, or a desktop), and, thus, are influenced by those contextual factors in their decision-making.
For instance, on a social media site, the appearance of some evocative content (\eg a controversial image or post) may elicit abusive behaviours from some users, which would not have occurred otherwise.
In fact, whenever there is a change on a website (\eg a new blog post or comment), this site also changes state, and, thus, the potential activity that will occur thereafter.
Such a change may, for example, affect the site's visibility on the Web (\eg search engine ranking), raise its profile (\eg Alexa ranking), or cause it to become a crime hotspot (\eg a target for raiding).
Likewise, the introduction of user input fields on a webpage (logins, search bars, contact forms -- possible site vulnerabilities) may be followed by SQL injection and XSS attacks, before which such attacks would not have been possible.
More generally, each time any online entity is updated, this entity changes state.
Thus, if one version of this entity has a vulnerability that is targeted by cybercriminals, whereas an updated version does not, these two versions would represent separate cyberplaces from the perspective of the malicious actors.
Therefore, users of unpatched versions of website and desktop software, for example, would be interacting with cyberplaces with elevated risks of victimisation (\eg software exploitation, drive-by download).
This component is relevant to both cyber-enabled and cyber-dependent crimes, as the states of websites and software are easily perceptible by both humans and machine actors, and could, therefore, influence their actions.

\subsubsection{The \textbf{Functional} Component.}
The third component of this cyberplace concept is the functional component, which encapsulates the intuitive function or purpose behind a cyberplace, to wit, why an actor (human or machine) interacts with it.
This is the cyberspace equivalent of the \emph{sense of place} aspect of physical places~\cite{cresswell2014place}, which refers to the feelings and emotions that are evoked when one considers a physical place (\eg a restaurant) as well as one's mental expectations concerning such a place (\eg the presence of tables, chairs, food and drink, and operating hours mainly within the afternoon or evening).
As such, this is the most abstract component of cyberplace.
However, it is also a vital one in that it provides the purposes to how and why actors interact (or expect to interact) with cyberplaces.
For instance, the main reason for accessing news websites is to acquire local news information.
News sites would probably be expected to categorise each news items with some (hyperlinked) headline, perhaps with some images or short synopses, and with some sort of order according to recency and/or significance (\eg recent and important news at the top of the page, older and less important news at the bottom or archived).
Generally, however, one would not expect to use a news website to purchase groceries, to buy a new laptop, or to access a remote server -- one would expect to visit an e-commerce site for the former scenarios, or to use a remote-desktop application for the latter.
In fact, it is probably for this reason that the Web is characterised using physical metaphors (sites, forums, chatrooms, e-mail, desktops, \etc) with which we are familiar in the real world.
Furthermore, machine actors also share an innate understanding of cyberplaces in accordance to their creators.
For instance, crawlers may be programmed to collect news items on a website with an inherent expectation of how those items are likely to be organised (\eg the HTML tags to look for within the source code).
Likewise, there are established protocols that enable computer systems to communicate with each other, depending on the services and ports involved (\eg HTTP/S and ports 80/443 for web content, S/FTP and ports 22/21 for file transfers).

In regards to cybercrime (and more generally, the way that the Internet is used), the functional component of cyberplaces is important in that it is likely the defining factor as to how and why human and machine actors interact with cyberplaces, as well as to where, when, and by whom they are used.
For example, as we discussed earlier, there is a general expectation that sites hosting illegal or explicit content (streaming, pirate, or pornographic sites) are more likely to host malware than others.
Yet, despite these risks, users are still drawn to these websites.
Thus, the functional aspect of such cyberplaces could be what makes them desirable vectors for malware distribution.

\subsection{Quantising Cyberplaces and Potential Applications} \label{sec:cyberplace_practice}

In a broad sense, we have defined the concept of cyberplace as a combination of three fundamental components.
Of course, we can (and wish to) examine this concept in further discourse: the internal relationships between these components; its relationships with the real-world and cyberspace concepts of space, time, place, people, and machine actors; how it maps to various computer system and telecommunication models; etc.
However, we will leave those discussions for future works and, for now, focus on how this concept may be applied in practice.

\subsubsection{Cyberplace Classification}
The three components of our cyberplace concept can be used to encapsulate all the information that is required to identify, describe, and differentiate cyberplaces on the Internet.
Any cyberplace can be described in terms of its \textit{function} (\eg server, computer, website, web application, process, file, and each with their own types), its \textit{location} (\eg IP address, MAC address, file directory, memory address), and its \textit{state}, which pertains to all tangible information relating to it at that point in time (\eg webpage or software source code, types of content, included libraries, software versions, active processes, open ports).
How one describes a cyberplace and to what level of detail would depend on the level of abstraction that is relevant to them and the underlying schema that they are applying.
For example, focusing on the case of web content (accompanied by an understanding of Web structure), one could use functional information to coarsely categorise individual websites as cyberplaces, then to categorise individual webpages as smaller cyberplaces, then to categorise the individual features on these pages (search bar, blog entries, video player, \etc) as even smaller ones, and so on and so forth.
This would, in turn, warrant the inclusion of finer-grain locational information (\eg from a domain and IP address, to the subdirectory of a webpage, to a line in the source code) and statal information (\eg from overall site attributes, such as the number of incoming or outgoing hyperlinks, or the server software version, to properties of individual pages or blocks of content, such as the number and content of images or HTML child elements, or the version of Javascript used).
Clearly, there are an infinite number of ways in which this process may be implemented.
Though such a classification could begin with manual efforts, a useful research direction may be to investigate the use of modern analytical techniques, such as through data science and machine learning, to heuristically extract the most efficient descriptors for cyberplace classification at each level of abstraction.

\subsubsection{Cyberplace Risk Modeling}
The next step in cybercrime analysis involves classifying cyberplaces into various types of cybercrime hotspots (see Section~\ref{sec:crime_patterns}) by applying the principles of crime pattern theory~\cite{brantingham_criminality_1995}.
For example, we could label e-commerce and CMS-enabled sites, and computers using the most popular operating systems, as potential crime generators (\ie they generate criminal opportunity due to the presence of many potential targets).
On the other hand, sites that serve illegal content (pirate and streaming sites, underground marketplaces) could be labelled as potential crime attractors (\ie they generate criminal opportunity as they are well-known for harbouring illegal activity).
Finally, some social media sites and apps, and sites with poor cyber hygiene in general could be labelled as potential crime enablers (\ie they generate criminal opportunity through lack of supervision or management).
Either by empirical analysis, by generating probabilistic models, or by some other method, labelling cyberplaces as cybercrime hotspots could further guide their classification by cyber risk.
Such classifications could express the likelihood of cybercriminal activity occurring at these cyberplaces, and, therefore, the risk of victimisation for their end-users and managers.
One could even go further to derive expected user activities and potential cybercriminal modus operandi by way of analysing how users interact with these cyberplaces (\eg UX analysis of websites and applications, control flow analysis of programs, vulnerability assessments, penetration testing).
Thus, these behavioural paths could be used as bases for establishing crime scripts for these cyberplaces (\ie sequences of events, decisions, and actions that cybercriminals may follow preceding and following a crime event).
Given that cyberspace is inherently \emph{data-rich} and \emph{discretised}, it is possible that these analytical models may be more suitable for this environment than the contiguous, real world.

\subsubsection{Cybercrime Mitigation}
Finally, these insights may be used for various mitigative strategies:

\noindent \textbf{Educating victims.}
Awareness campaigns could be raised concerning the types of sites, services, and applications that pose the greatest risk of harm towards users or particular demographics, with the potential for new software to provide warnings before entering or using a high-risk site or application.
In the same vein, providers of these cyberplaces (\eg website owners, app developers) could be informed of their likely cyber risk level, and how they could minimise these risks.

\noindent \textbf{Altering cyberplace characteristics.}
Another class of mitigations involves altering these cyberplaces to reduce criminal opportunity.
In line with various situational crime prevention approaches, this could range from hardening these cyberplaces against vulnerability exploitation and improving security hygiene, to altering UX features and control flows of websites and applications in order to minimise malevolent or criminal opportunity (\eg disabling or correcting vulnerable components that are likely to be exploited, requiring authentication in order to access a website).

\noindent \textbf{Applying tools and mitigation efforts more efficiently.}
As research on repeat victimisation suggests, the Pareto principle applies in that a majority of crimes only involves a minority of offenders and victims.
This suggests that the greatest reductions in cybercrime will arise by focusing cybersecurity tools and mitigative efforts towards educating and protecting the most exploited victims and targets, making the most crime-facilitating cyberplaces safer, and deterring, detecting, and apprehending the most prolific cybercriminals.
Cyberplace classification techniques enable such efforts by identifying the cybercrime hotspots on the Internet, the cybercriminals who operate in or target them, and the users who are victimised there.


\section{Related Work}
\label{ref:related}




\subsection{Environmental Criminology and Cybercrime}
In recent years there has been a steady increase in scientific studies seeking to use (and evaluate) environmental criminology theories in order to explain various forms of cybercrime, or whether it is a novel form of crime requiring new theories.
For almost two decades, discussions have been ongoing on the potential (multiplying) effect that digitalisation has had on crime \cite{grabosky_virtual_2001,wall2007cybercrime}.
Grabosky \cite{grabosky_virtual_2001} reflects on these discussions, concluding that, though the motivations behind crime and human nature are still the same, technology has enabled an increase and diversification in criminal opportunities through anonymising technologies, transatlantic connectivity, and an absence of clear-cut boundaries for potential guardianship.
Paoli~\cite{de2018engineer} argues that a sociotechnical perspective can be developed to understand this ``Novelty of Cybercrime''
using some insights from criminology.
In particular,
Paoli iterates that ``engineer-criminologists'' (unknowingly) employed vocabulary, concepts, and techniques very similar to those found in criminology to mitigate the new threat of malicious users in timesharing systems in the 1960s and 70s.
Although environmental criminology has made a steady focal shift into cybercrime with studies assessing its suitability for this form of crime \cite{leukfeldt_applying_2016,holt_examining_2008,yar_novelty_2005}, formal approaches to conceptualising `places' in cyberspace (or `cyberplaces') in relation to cybercrime are almost non-existent.
Llinares and Johnson \cite{llinares_cybercrime_nodate} review the applicability of environmental criminology to crimes in cyberspace, particularly evaluating the virtual places of cybercrime and how they differ from their physical counterparts.
However, the lack of development of a `cyberplace' concept is a critical gap in research, as `place' is fundamental to the entirety of environmental criminology and its practices.
In this work, we go much further than prior studies in providing an overview of cybercrime research from two disciplinary perspectives: information security and environmental criminology.
We draw parallels between these two understandings of cybercrime, highlight areas of overlap, and reason that future works should utilise these complementary approaches for a holistic approach towards cybercrime prevention.
We initiate this process in our conceptualisation of \emph{cyberplace}, and we propose how this concept could be applied for cybercrime analysis and prevention using environmental criminology theories and practices.

\subsection{Concepts of `Place' in Cyberspace.}
Researchers and professionals in a variety of fields have made concerted attempts to formally establish a concept of `cyberplaces', or virtual locations.
In the geographical sciences, Tranos and Nijkamp~\cite{tranos_death_2013} study the impact of physical distance on the formation of the Internet infrastructure, and whether physical distance survives in virtual geography, even after controlling for relational proximities.
On the other hand, in the field of urban technology, Devriendt \etal~\cite{devriendt_cyberplace_2008} identify two approaches to analysing ``virtual'' or digital intercity linkages (\ie linkages based on ICT).
In both of these works, they utilise the same geographic metaphors of \emph{cyber-place} (CP), which is defined as the projection of the infrastructural layer of cyberspace on traditional space, and \emph{cyberspace} (CS), which is defined as the virtual or immaterial world wherein people communicate with each other via networked technologies, and that physical laws and aspects, such as distance and time, are practically irrelevant.
From a sociological perspective, Wellman~\cite{wellman_computer_2001,wellman_physical_2001} characterises computer networks as social networks, and thus argues that they should not be studied in isolation, but as integral parts of everyday life.
An example of such studies includes the work of Sussan \etal~\cite{sussan2006location} on how cyberspace allows consumers to form virtual communities and engage in online word-of-mouth exchanges.
Wellman~\cite{wellman_physical_2001} initially thinks of computer-to-computer interactions becoming increasingly ``placeless''.
Nonetheless, in reference to the development of place-based social networks, the author refers to ``online relationships and communities'' being ``truly in cyberplaces, and not just cyberspaces'', alluding to cyberplaces as online services that enable peer-to-peer activity.
Significant efforts have also been made in the legal sector in isolating a licit definition of `place' in cyberspace.
Hunter~\cite{hunter_cyberspace_2003} discusses the \textsc{cyberspace as place} legal metaphor, which was commonly used in the U.S. to understand Internet communication as ``having certain spatial characteristics from our physical world experience'', thus giving legal precedent in cases involving Internet services and digital property.
Lemley~\cite{lemley_place_2003} argues that the Internet is dominated by publicly accessible sites or spaces, so law should not assume every part of cyberspace is ``owned'' by a particular entity.
The author also contradicts the \textsc{cyberspace as place} metaphor, mainly due to large disparities between the physical idea of property and the cyber world.
However, the author does not discuss the use of synchronous applications, such as social networks or instant messengers, 
and how they relate to this metaphor.
The author also fails to consider individual websites as places, which may serve as a better analogy.
Cohen and Hiller~\cite{cohen_towards_2003} discuss the legal definition of `place' (U.S. law) and attempt to define an analogous counterpart for `cyberplace' for the purposes of clarifying the relevant laws and rights.
In particular, they note that the \textsc{cyberspace as place}
metaphor is far too broad a definition, and suggest a new framework that identifies when a private provider of online content or access creates a `place of public communication,' with the purpose of disambiguating
between private and public spaces on the Internet, much like in the physical world, for conflict resolution.
Our definition of cyberplace significantly differs to that of prior works in that we derive a holistic concept of cyberplace, which takes into account cyberspace, online activity (both benign and malicious), and their relationships with the real world.
In a sense, our concept combines the CP and CS metaphors of urban technology, while remaining applicable to any relevant field.



\section{Conclusion}
\label{sec:conclusion}

In this paper, we conducted a literature review of cybercrime research from the perspectives of information security and environmental criminology.
We presented an overview of how these two fields understand and (could) deal with cybercrime, identifying connections between their apparently disparate approaches.
Upon review of a wide array of literature and cybercrime contexts, we provide motivating evidence as to why a new, complementary research approach should be pursued involving these two fields.
We initiate this process in earnest, first, by showing how frameworks from environmental criminology could be used to devise new cybercrime countermeasures; second, by proposing a conceptualisation of the immediate environmental contexts (or \textit{cyberplaces}) where cybercrimes occur; and third, by providing some motivating examples of how the concept of cyberplaces, together with environmental criminology, could be used to better analyse and mitigate cybercrime.
We hope that this work will encourage the wider research community to build upon this cyberplace concept and its implementation in the transfer of crime prevention theories and frameworks between environmental criminology and information security.
Above all, we hope that such collaborations will yield new and better approaches to cybercrime prevention and provide systematic frameworks that inform practitioners on the full range of techniques available.

\begin{acks}

We would like to thank all those who reviewed this work.
Colin C. Ife is supported by the Dawes Centre for Future Crimes, and EPSRC under grant EP/M507970/1.
Steven J. Murdoch is supported by The Royal Society under grant UF160505.

\end{acks}

\bibliographystyle{ACM-Reference-Format}
\bibliography{biblio,env_crim,cyb_pl,cyb_ex}

\end{document}